\documentclass[aps,pra,amsmath,amssymb,superscriptaddress,twocolumn,notitlepage,nofootinbib,longbibliography]{revtex4-2}

\usepackage{times,bbm,amsmath,amssymb,amsthm}
\usepackage{braket}
\usepackage{epsfig,color}
\usepackage{xcolor}
\usepackage[colorlinks=true,citecolor=blue,linkcolor=blue]{hyperref}
\hypersetup{
    allcolors  = {blue},
}
\usepackage{subfigure}
\usepackage{graphicx}
\usepackage{algorithm}
\usepackage{algpseudocode}
\usepackage{xpatch} 
\usepackage{xspace}
\usepackage{float}
\usepackage{placeins}
\usepackage[toc,page]{appendix}

\makeatletter
\xpatchcmd\@collaboration@present{(}{\medskip}{}{}
\xpatchcmd\@collaboration@present{)}{}{}{}
\makeatother

\newcommand{\parti}[0]{\mathbf{\mathcal{K}}}
\newcommand{\setparti}[0]{\mathcal{K}}  
\newcommand{\sizeparti}[0]{K}  

\newcommand{\vect}[1]{\boldsymbol{#1}}

\newcommand{\op}[1]{\hat{#1}}
\newcommand{\perm}[1]{\text{perm}\left({#1}\right)}

\newcommand{\valery}{Shchesnovich \xspace}
\DeclareMathOperator*{\E}{\mathbb{E}}

\definecolor{chromeyellow}{rgb}{1.0, 0.65, 0.0}

\begin{document}
\title{\textsc{Experimental validation of boson sampling using detector binning}}

\begin{abstract}
    
    We experimentally demonstrate a testing strategy for boson samplers that is based on efficiently computable expressions for the output photon counting distributions binned over multiple optical modes. We apply this method to validate boson sampling experiments with three photons on a reconfigurable photonic chip, which implements a four-mode interferometer, analyzing 50 Haar-random unitary transformations while tuning photon distinguishability via controlled delays. 
    We show that for high values of indistinguishability, the experiment accurately reproduces the ideal boson sampling binned-mode distributions, which exhibit variations that depend both on the specific interferometer implemented as well as the choice of bin, confirming the usefulness of the method to diagnose imperfections such as partial distinguishability or imperfect chip control. Finally, we analyze the behavior of Haar-averaged binned-mode distributions with partial distinguishability and demonstrate analytically that its variance is proportional to the average of the square of the photons' indistinguishability parameter. These findings highlight the central role of binning in boson sampling validation, offering a scalable and efficient framework for assessing multiphoton interference and experimental performance.
\end{abstract}

\author{Malaquias Correa Anguita}
\thanks{These two authors contributed equally}
\email{m.correaanguita@utwente.nl}
\affiliation{MESA+ Institute for Nanotechnology, University of Twente, 7522 NB Enschede, The Netherlands}

\author{Anita Camillini}
\thanks{These two authors contributed equally}
\email{a.camillini@cineca.it}
\affiliation{International Iberian Nanotechnology Laboratory (INL) Av. Mestre José Veiga s/n, 4715-330 Braga, Portugal}
\affiliation{CINECA Consorzio Interuniversitario, Via Magnanelli 6/3, 40033 Casalecchio di Reno, Italy}
\affiliation{Centro de Física, Universidade do Minho, Braga 4710-057, Portugal}

\author{Sara Marzban}
\affiliation{MESA+ Institute for Nanotechnology, University of Twente, 7522 NB Enschede, The Netherlands}

\author{Marco Robbio}
\affiliation{International Iberian Nanotechnology Laboratory (INL) Av. Mestre José Veiga s/n, 4715-330 Braga, Portugal}
\affiliation{Quantum Information and Communication, Ecole polytechnique de Bruxelles, CP 165/59, Université libre de Bruxelles (ULB),
1050 Brussels, Belgium}

\author{Benoit Seron}
\affiliation{Quantum Information and Communication, Ecole polytechnique de Bruxelles, CP 165/59, Université libre de Bruxelles (ULB),
1050 Brussels, Belgium}
\affiliation{Physikalisches Institut, Albert-Ludwigs-Universit\"at Freiburg,
Hermann-Herder-Stra{\ss}e 3, D-79104 Freiburg, Germany}
\affiliation{EUCOR Centre for Quantum Science and Quantum Computing,
Albert-Ludwigs-Universit\"at Freiburg, Hermann-Herder-Stra{\ss}e 3, D-79104 Freiburg, Germany}
\author{Leonardo Novo}
\affiliation{International Iberian Nanotechnology Laboratory (INL) Av. Mestre José Veiga s/n, 4715-330 Braga, Portugal}

\author{Jelmer J. Renema}
\affiliation{MESA+ Institute for Nanotechnology, University of Twente, 7522 NB Enschede, The Netherlands}


\maketitle

\section{Introduction}
\label{sec:Intro}
An intermediate goal in building a full-scale quantum computer is the demonstration that certain well-defined computational tasks can be efficiently carried out by quantum hardware while being out of reach for even the best classical supercomputers. Recent experimental breakthrough works have claimed that this major goal has been achieved, both in quantum optical devices \cite{zhong2020science, madsen2022quantum, zhong2021phase, deng2023gaussian} as well as other platforms \cite{arute2019quantum, wu2021strong, gao2024establishing}. 

A seminal work that inspired many of these developments was the introduction of the boson sampling problem by Aaronson and Arkhipov \cite{aaronson2011computational}. This work laid out the theoretical foundations used to support claims of quantum computational advantage for sampling problems. The problem consists of sending non-interacting bosons (usually single photons) through a linear multimode interferometer which are then detected at the output of the device (see Fig.~\ref{fig:scheme_verification}). At the heart of the computational complexity of the boson sampling problem lies the fact that the different outcome probabilities are proportional to matrix permanents, which take exponential time to compute with classical computers. While simple in its formulation, boson sampling provides an accessible platform to disprove the extended Church-Turing thesis.  

As boson sampling experiments grow in size, it is of importance that scalable methods exist for checking that the device is working correctly, in order to substantiate claims of a quantum advantage. It is known that, if we only have access to the output data from the device and the specifications of the sampling problem, it is impossible to efficiently certify without any loopholes that the device is indeed approximately sampling from the desired distribution \cite{hangleiter_sample_complexity, shchesnovich2022boson,villalonga2021efficient}. However, a large amount of work has been devoted to the development of validation tests, where the aim is to distinguish an ideal boson sampling experiment from a noisy one or else from certain mock-up samplers that can be efficiently implemented using classical computers \cite{renema2018efficient,oszmaniec2018classical,garcia2019simulating,brod2020classical,renema2018classical, mezher2022assessing}. 

In the context of validation, methods based on binning the outcome distribution, by jointly counting the photons in groups of detectors, have been recently developed, both for single-photon boson sampling~\cite{seron2022efficient} and Gaussian boson sampling \cite{bressanini2023gaussian, opanchuk2018simulating, drummond2022simulating, dellios2022validation, bulmer2024simulating}. Some advantages of these methods are that they naturally generalize previous validation tests based on marginal distributions \cite{villalonga2021efficient} (or equivalently, multimode correlators \cite{walschaers_statistical_2016, phillips_benchmarking_2019}) and generalized bunching probabilities \cite{shchesnovich2016universality,shchesnovich2021distinguishing}, which can be recovered as special cases of this framework. At the same time, it has been shown that there are efficient ways to approximately predict the binned distribution resulting from an ideal boson sampler independently of the size of the bins, as long as the total number of bins is fixed \cite{seron2022efficient}. Hence, a comparison between the theoretically predicted binned distribution and the experimentally measured one for random bin choices, provides a simple way to obtain a lower bound of the distance between the ideal and experimental boson sampling outcome distribution. This is in contrast with other ideas for validation based on heavy output generation \cite{martinez2024linear, hangleiter2023computational, oh2023spoofing} or Bayesian tests, which require exponentially hard classical computations \cite{bentivegna2015bayesian, renema2021}.  To our knowledge, no efficient classical algorithm exists that is able to pass the validation test based on binned-mode photon number distributions. The difficulty of spoofing binned-mode distributions plays a central role in the recently proposed application of boson sampling as a quantum proof-of-work scheme for blockchain consensus \cite{Singh2025_Pow_BS}. In this scheme, binned-mode distributions are used for validation purposes, whereas other binning strategies of the state space -- for which there is currently no efficient classical simulation scheme -- are used to implement cryptographic one-way functions \cite{nikolopoulos2016decision, nikolopoulos2019cryptographic, wang2023_exp_crypto}.  

In this work, we apply binned-mode photon number distributions to experimentally validate three-photon boson sampling experiments on a reconfigurable photonic chip. We implement 50 Haar random interferometers and analyze the output data according to several different bin choices, obtaining good agreement with the theoretically predicted values. To test the sensitivity of the method to common sources of experimental errors, we tune the distinguishability between the input photons via controlled time delays, showing that binned distributions are highly sensitive to partial distinguishability. 

Moreover, we extend the theory by deriving an analytic expression for the variance of the Haar-averaged binned-mode distribution of a single bin, showing that it is directly proportional to the average Hong-Ou-Mandel visibilities of the input photons. This explains the behavior that was numerically predicted in \cite{seron2022efficient} and experimentally observed in the present work: Haar-averaged binned photon number distributions tend to get wider as photons become more and more indistinguishable, a clear signature of the tendency for photons to bunch. Hence, the experimental estimation of the variance of the binned-mode distribution yields a practical scheme to obtain guarantees on photonic indistinguishability by a simple postprocessing of boson sampling data, extending previous work connecting single-mode photon number variances to photonic indistinguishability \cite{rodari_experimental_2024}.   

Overall, these results provide an experimental demonstration of the centrality of binned distributions to boson sampling, from the point of view of potential applications, validation, and the theory of multiphoton interference.


\section{Preliminaries}
\label{sec:prelim}

\begin{figure}[t]
    \centering
    \includegraphics[width=0.45\textwidth]{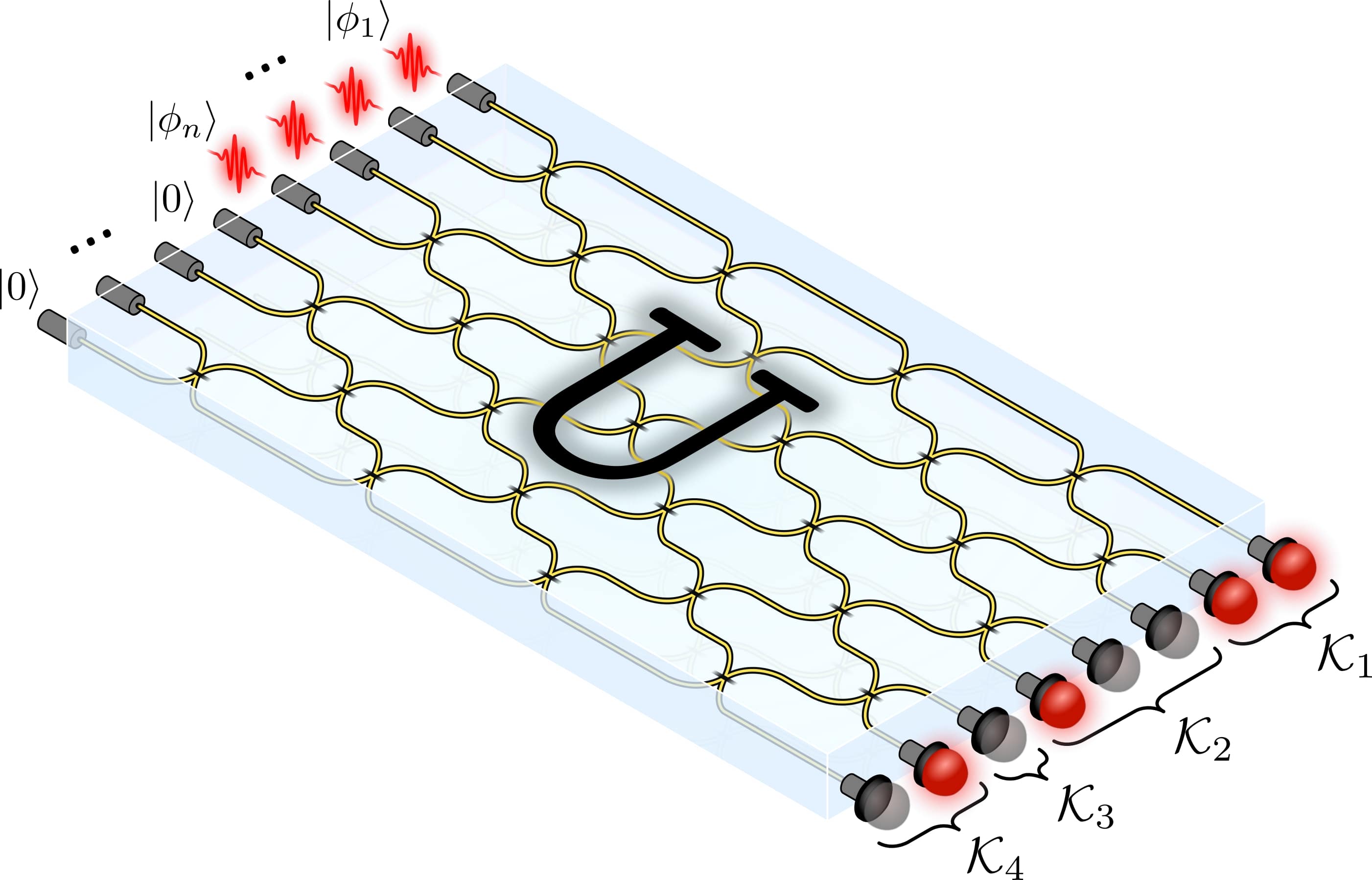}
    \caption{A boson sampling experiment where $n$ single photons at the input, with possibly different internal wave-functions $\ket{\phi_i}$, are sent through the first $n$ input modes of a linear interferometer $U$. The validation test discussed in this work is based on coarse-graining the boson sampling distribution by grouping the output modes into a few bins. In the figure, we represent in red the detectors that have detected one photon and in gray the detectors that did not click. In this example, we have 4 bins and the binned outcome observed is $\mathbf{k} = (2, 1, 0, 1)$.}
    \label{fig:scheme_verification}
\end{figure}

In this section, we provide an overview of the methodology used to compute binned distributions of noisy and ideal boson samplers from \cite{seron2022efficient}. At first, we discuss the primary imperfections affecting boson sampling experiments, focusing on the modeling of partial distinguishability (Sec.~\ref{sec:Prem-NoisyBS}). 
Next, we introduce the concept of binned-mode photon-number distributions. The binning approach consists of grouping into bins the photon number resolving detectors present at the output of each mode of the interferometer, as shown in Fig. \ref{fig:scheme_verification}. In Sec~\ref{sec:Prelim-binnedDist}, we summarize how to efficiently compute the probability of finding a given number of photons in each bin, when the number of bins is fixed, which can be used for comparison with experimental data for validation purposes. 

\subsection{Noisy Boson Sampling}
\label{sec:Prem-NoisyBS}
The main imperfections affecting boson sampling experiments are losses and partial distinguishability between the photons. The latter can be due to small time delays between the photons arriving in different input modes, or differences in internal degrees of freedom such as spectral distribution or polarization. Although the effects of photon losses can be partially corrected by postselecting on outcomes where the number of photons at the output matches the input, this does create second order errors whose impact is negligible and can be reasonably disregarded. In contrast, as photon detectors are usually insensitive to degrees of freedom like arrival time or spectral shape, it is generally not possible to correct for partial distinguishability via postselection. In this work, we confine our study to distinguishability, as this imperfection dampens multiphoton interference effects, which can be exploited by classical simulation algorithms, and thus lead to an experiment that is efficient to simulate classically \cite{renema2018efficient, moylett2019classically, hoven2024}. 

Following previous works \cite{tichy2015_partial_distinguishability, shchesnovich2015partial, shchesnovich2021distinguishing, shchesnovich2017partial, shchesnovich2015partial}, we model partial distinguishability by describing each photon through its spatial input mode $i$ as well as other degrees of freedom encompassed in an internal wave function $\ket{\phi_i}$. We denote the creation operator of the photon at the $i$th spatial input mode as $\op{a}_{i, \phi_i}^{\dagger}$. The operator $\hat{U}$, representing the linear interferometer, can be fully described in terms of an $m\times m$ unitary matrix $U$, via its action on the creation operators 
\begin{equation}
    \op{U} \op{a}_{j, \phi_j}^\dagger \op{U}^\dagger = \sum_{k=1}^m U_{k,j} \op{a}_{k, \phi_j}^\dagger, 
\end{equation}
which acts on spatial degrees of freedom while leaving internal degrees of freedom invariant. Assuming that the detectors count the number of photons in a given spatial output mode (independently of the internal state), the probability of observing a given output photo-counting distribution depends only on the matrix $U$ and on the \textit{Gram} matrix $\mathcal{X}$, whose entries are defined by the overlaps between the internal wavefunctions of the different input photons
\begin{equation}
x_{ij} = \braket{\phi_i|\phi_j}.
\end{equation}
The general expression to compute the probabilitiy of a given outcome pattern $(s_1, ..., s_m)$, where $s_j$ denotes the number of photons detected by the $j$th detector, involves the evaluation of a tensor permanent, which reduces to a simple matrix permanent in the ideal case of fully indistinguishable photons \cite{tichy2015_partial_distinguishability}. In this work, however, we do not focus on individual outcome probabilities, but instead on binned distributions obtained by jointly counting the total number of photos in different groups of detectors.  

\subsection{Binned-mode photon-number distributions}
\label{sec:Prelim-binnedDist}
Consider a boson sampling experiment with $n$ photons in $m>n$ modes. There are exponentially many possible outcomes, each one happening with a probability that is typically exponentially small in $n$. This implies that a good approximation of individual outcome probabilities cannot be done with a polynomial number of samples. For this reason it is interesting to coarse-grain the space of possible outcomes. The approach we adopt in this work is to group together the output modes into different bins, and jointly count the number of photons arriving in each bin. For a constant number of bins $K$, the binned photo-counting distribution has only polynomially many outcomes, and so a meaningful approximation of this probability distribution can be done with polynomially many samples.

In this section, we summarize an efficient method for computing such binned-mode outcome distributions proposed in \cite{seron2022efficient} and introduce the notations and terminology used in the rest of the article. Let us consider the partition $\parti=\{\setparti_1,\dots,\setparti_\sizeparti\}$ of the output modes $\mathcal{M}=\{1,2,\dots, m\}$ of the interferometer into $\sizeparti$ nonempty and mutually disjoint subsets $\setparti_z\subset \mathcal{M}$, with $z\in \{1,\dots,\sizeparti\}$, which define the different bins. 
If the photons' configuration at the output of the boson sampler is $\vect{s}= (s_1, s_2, \dots, s_m)$, the outcome after the binning is defined by a vector $\vect{k}$ of dimension $\sizeparti$, whose components are given by  
\begin{equation}
 k_z= \sum_{j\in \setparti_z} s_j .
\end{equation}
The set of possible binned outcomes is defined as 
\begin{equation}
    \Omega^{\sizeparti}= \{(k_1, k_2, \dots, k_{\sizeparti})~|~ k_z\in \Omega, \forall z\in \{1,\dots,\sizeparti\}\},
\end{equation}
with $\Omega= \{0,1,\dots,n\}$. 
Moreover, the number of photons in a given bin is associated to the observable
\begin{equation}
    \op{N}_{\setparti_z} = \sum _{j \in \setparti_z} \op{n}_j,
\end{equation}
where $\op{n}_j$ is the photon number operator of a given spatial mode. 

We are interested in computing the probabilities $P(\vect{k})$ of observing the different possible photon number configurations in this partition. It is clear that computing each probability $P(\vect{k})$ by summing the individual probabilities of events $(s_1,...., s_m)$ that contribute to a binned outcome $\vect{k}$ is inefficient, as there are in general exponentially many outcomes contributing to this probability. A more efficient method is to compute the distribution via its characteristic function
\begin{align}
    \label{eq:characteristicFunctionDef}
    x({\vect{\eta}})&= 
    \E_{\vect{k}}\left[ \exp\left(i  \vect{\eta}\cdot \vect{k}\right)\right]\\
    & = \sum_{\vect{k}\in \Omega^{K}}P({\vect{k}}) \exp\left( i \vect{\eta}\cdot \vect{k}\right), 
\end{align}
with ${\vect{\eta}} \in \mathbb{R}^K$.
For a given initial state of partially distinguishable photons $\ket{\Psi_{in}}$, 
it can be shown that the characteristic function $x(\vect{\eta})$ can be computed as the quantum expectation value
\begin{align}
    x({\vect{\eta}})&= \bra{\Psi_\mathrm{out}}e^{i \vect{\eta}\cdot \vect{\hat{N}_{\mathcal{K}}}} \ket{\Psi_\mathrm{out}}\\
    ~&=\bra{\Psi_\mathrm{in}} \hat{U}^\dagger e^{i \vect{\eta}\cdot \vect{\hat{N}_{\mathcal{K}}}} \hat{U}\ket{\Psi_\mathrm{in}}\label{eq:characteristic_interferometer}, 
\end{align}
where we have used the notation
\begin{equation}
    \vect{\eta}\cdot \vect{\hat{N}_{\mathcal{K}}}= \sum_{z=1}^{\sizeparti} \eta_z \op{N}_{\setparti_z}.
\end{equation}
This can be seen as a particular transition amplitude of an interferometric process with interferometer $\op{V}(\vect{\eta})=\op{U}^{\dagger} e^{i \vect{\eta}\cdot \vect{\hat{N}_{\mathcal{K}}}} 
\op{U}$, which is given in terms of a matrix permanent (see Appendix~\ref{sec:app_charfunction_permanent}).  This matrix permanent can either be evaluated exactly in time $\mathcal{O}(n 2^n )$ or up to an additive error $\epsilon$ in time $\mathcal{O}(n^2/\epsilon^2)$ via Gurvits' randomized algorithm for the approximation of matrix permanents. This way, the binned-mode outcome probability distribution $P(\vect{k})$ can be retrieved by evaluating $x({\vect{\eta}})$ at $(n+1)^{\sizeparti}$ points on a $\sizeparti$-dimensional grid, namely,
\begin{align}
\vect{\nu_{\vect{l}}} = \frac{2 \pi  \vect{l}}{n+1} ,
\quad \text{with~} l_z\in \Omega, \forall z\in \{1,\dots,\sizeparti\}
\end{align}
and taking the multidimensional Fourier transform  
\begin{align}\label{eq:Pk_from_xl}
    P(\vect{k})= \frac{1}{(n+1)^K}\sum_{\vect{l}\in \Omega^{\sizeparti}} x(\vect{\nu_{\vect{l}}}) \exp\left(-i \vect{\nu_{\vect{l}}} \cdot \vect{k}\right). 
\end{align}
It is shown in \cite{seron2022efficient} that this method allows for the distribution $P(\vect{k})$ to be computed exactly in time 
\begin{equation}
    T_{\text{exact}} = \mathcal{O}\left(\sizeparti n(n+1)^\sizeparti \log(n+1)  2^n\right).
\end{equation}
for any given interferometric scenario. Moreover, it is possible to approximate the probabilities $P(\vect{k})$,
up to total variation distance $\beta$, \emph{in polynomial time} 
\begin{equation}
    T_{\text{approx}} = \mathcal{O}\left(n^{2K+2}\log(n) \beta^{-2}\right)
\end{equation}
given some theoretical model for the experiment which may include partial distinguishability between the photons as well as losses. 
In summary, independently of the size of each bin, as long as the number of bins is constant, i.e. it does not grow with the number of photons,  it is possible to efficiently estimate binned-mode distributions, even for large experiments.
A similar outcome was obtained for Gaussian Boson Sampling in \cite{bressanini2023gaussian}, although in this scenario the scaling of the approximation algorithm is in terms of the additive error is exponentially better. Alternative techniques to compute binned-mode outcome distributions of Gaussian or Fock state boson samplers can be found in Refs.~\cite{opanchuk2018simulating, drummond2022simulating, dellios2022validation, bulmer2024simulating}. 

Finally, it is interesting to note that binned-mode probability distributions encompass, as a particular case, marginal distributions over $K$ modes --  in this case, each bin would contain a single mode -- which have often been used in the context of boson sampling validation \cite{villalonga2021efficient, van2021experimental, giordani2018experimental}. Moreover, a \textit{generalized bunching probability} (GBP), defined as the probability that all $n$ photons are observed in a given subset of the output modes, is simply a particular outcome of a binned-mode distribution with a single bin. These probabilities are also interesting in the context of boson sampling validation \cite{shchesnovich2016universality,shchesnovich2021distinguishing, seron2022efficient, bressanini2023gaussian, mezher2022assessing, young2023atomic, opanchuk2018simulating, drummond2022simulating, dellios2022validation}. Indeed, it was argued that generalized bunching phenomena are sensitive to genuine multiphoton interference \cite{shchesnovich2021distinguishing}, contrary to few-mode marginals. Hence, in most scenarios, partial distinguishability will decrease these probabilities. We discuss this method in more detail in Section ~\ref{sec:theory-GBP}, as we also build a test for our experimental apparatus by measuring GBPs.

\section{Boson sampling validation with binned-mode distributions}
\label{sec:theory}

In this section, we delve into the validation of boson sampling experiments using binned-mode distributions. This method generalises and expands a variety of statistical validation tests for boson samplers, capturing bunching phenomena and providing a robust framework for validating experimental data. 


Finally, we explore the Haar-averaged behaviour of binned-mode photon number distributions over all possible interferometers, showing that its variance can be used as a good indicator of photonic indistinguishability. 

\subsection{Binned-modes validation test}
\label{sec:theory-validation}

In the context of validation tests, the purpose is not to fully certify that the experimental boson sampling device is sampling from the ideal distribution, as this would not be possible in an efficient manner \cite{hangleiter_sample_complexity}. Instead, the idea is to be able to discard a "bad" boson sampler, by showing that a noisy device samples from a distribution that is far from the ideal one. 

The binned-mode validation test allows us to easily access a lower bound on the Total Variation Distance (TVD) between the probability distribution sampled via the experiment and the one from an ideal boson sampler. Defining $p(\vect{s})$ as the probability of each photon's configuration $\vect{s}$ at the detectors' output, the TVD between experimental data $p^{\text{exp}} = \{ p^{\text{exp}}(\vect{s}) \}$ and the corresponding theoretical Boson Sampling distribution $p^{\text{th}} = \{ p^{\text{th}}(\vect{s}) \}$ is given by
\begin{equation}
{TVD}^{\text{full}} = \frac{1}{2} \sum_{\vect{s}} \left| p^{\text{exp}}(\vect{s}) - p^{\text{th}}(\vect{s}) \right|
\end{equation}
As neither $p^{\text{th}}(\vect{s})$  can be efficiently calculated, nor $p^{\text{exp}}(\vect{s})$ can be efficiently estimated experimentally, it is useful to look at the binned distribution for some predefined choice of bins $P^{\text{exp}} = \{ P^{\text{exp}}(\vect{k}) \}$ and $P^{\text{th}} = \{ P^{\text{th}}(\vect{k}) \}$. It is possible to see that the TVD between theoretical and experimental binned distribution
\begin{equation}
    \label{eq.tvd-bin}
    TVD^{\text{binned}} = \frac{1}{2} \sum_{\vect{k}} \left| P^{\text{exp}}(\vect{k}) - P^{\text{th}}(\vect{k}) \right|, 
\end{equation}
where the sum runs over all $\vect{k} = (k_1, \dots, k_\sizeparti)$ such that $\sum k_i = n$, gives a lower bound to the TVD between the full distributions, i.e. 
\begin{equation}
    TVD^{\text{full}} \geq TVD^{\text{binned}}.
\end{equation}
This is a direct consequence of the fact that many outcomes $\vect{s}$ will contribute to the same binned-mode outcome $\vect{k}$. 
Given that this result holds for any partition choice, the same experimental data can be analyzed through different partitions and the worst-case $TVD^{\text{binned}}$  can be taken as the lower bound to $TVD^{\text{full}}$.

A few remarks are in order. The choice of the number of bins $K$ will depend on the classical computational power, since the complexity of approximating the ideal boson sampling distribution scales as $n^{2K+2}$, but also on the number of experimentally available samples, to accurately estimate the values $P^{\text{exp}}(\vect{k})$. As shown numerically in \cite{seron2022efficient} and experimentally in this work (see Sec.~\ref{sec:res-val}), looking at the photon number distribution in a single bin seems to provide a good way to distinguish an ideal boson sampler for one with sufficiently partially distinguishable photons. This claim is also supported by theoretical work arguing that generalized bunching probabilities are sensitive to all orders of multiphoton interference \cite{shchesnovich2016universality, shchesnovich2021distinguishing}.    

It is also worth noting that if one has a theoretical model that describes the experiment, for example a boson sampler with partially distinguishable photons, one may use binned-mode distribution to validate this theoretical modeling of the experiment. If significant deviations are found, then one may conclude that other noise sources are at play.

Finally we point out that, for a given number of bins $\sizeparti$, there exist 
\begin{equation}
    C = \binom{m}{\setparti_1, \dots, \setparti_\sizeparti}
\end{equation}
different ways of grouping detectors. The binned-mode distribution varies for different choices (see Appendix C of \cite{seron2022efficient}). This makes it challenging for any efficient classical mock-up sampler to pass binned-mode distribution tests without sampling for the ideal boson sampling distribution. To our knowledge, there is currently no classical algorithm that is able to pass this test, contrary to other validation tests based on marginals \cite{villalonga2021efficient}.

\subsubsection{Generalised bunching probabilities}
\label{sec:theory-GBP}
In an ideal boson sampler with indistinguishable bosons, bunching phenomena are expected be more prominent than in an imperfect boson sampler with partially distinguishable particles \cite{shchesnovich2016universality, shchesnovich2016permanent, seronBosonBunching, pioge2023enhanced}. Binned-mode outcome probability distribution capture such generalized bunching phenomena, namely, the tendency for bosons to bunch in any given subset of the output modes. Following Shchesnovich \cite{shchesnovich2016universality}, we can define the \emph{generalized bunching probability} $P_n(\mathcal{K})$, of finding all $n$ photons in a given subset $\mathcal{K}$ of the output modes, which is a particular outcome probability of a binned-mode distribution. For a given distinguishability matrix $\mathcal{X}$, this probability is given by 
\begin{equation}
\label{eq:bunching_probability}
P_{n}(\mathcal{X}) = \text{Perm}(H \odot \mathcal{X}),
\end{equation}
where Perm${(\cdot)}$ denotes the matrix permanent while $\odot$ denotes the Hadamard (or element-wise) product, that is,  $(H \odot \mathcal{X})_{i,j} = H_{i,j} \,\mathcal{X}_{i,j}$ and where the matrix $H$ is defined as 
\begin{equation}
\label{eq:H_Shchesnovich}
H_{i,j} = \sum_{k \in \mathcal{K}} U_{k,i}^* U_{k,j}, \qquad i,j\in \{1,...,n\}. 
\end{equation}
For three-photon experiments, it can be proved that $P_{n}(\mathcal{X})$ is always maximized if bosons are fully indistinguishable, i.e. when $x_{i,j}=1 \forall i,j $, for any interferometer and any choice of subset \cite{shchesnovich2016universality}. A few exceptions are known to this behavior for larger photon numbers but they require specific choices of the interferometer as well as of the Gram matrix characterizing the distinguishability between the photons \cite{seronBosonBunching, pioge2023enhanced}. Numerical evidence suggests that, in the vast majority of cases, partial distinguishability decreases generalized bunching probabilities, and so the estimation of generalized bunching probabilities and respective comparison to the expected ones from an ideal boson sampler, is a good method to validate whether the device is working correctly \cite{shchesnovich2016universality, seronBosonBunching}.


\begin{figure*}[t]
    \centering
    \includegraphics[width = 0.95\textwidth]{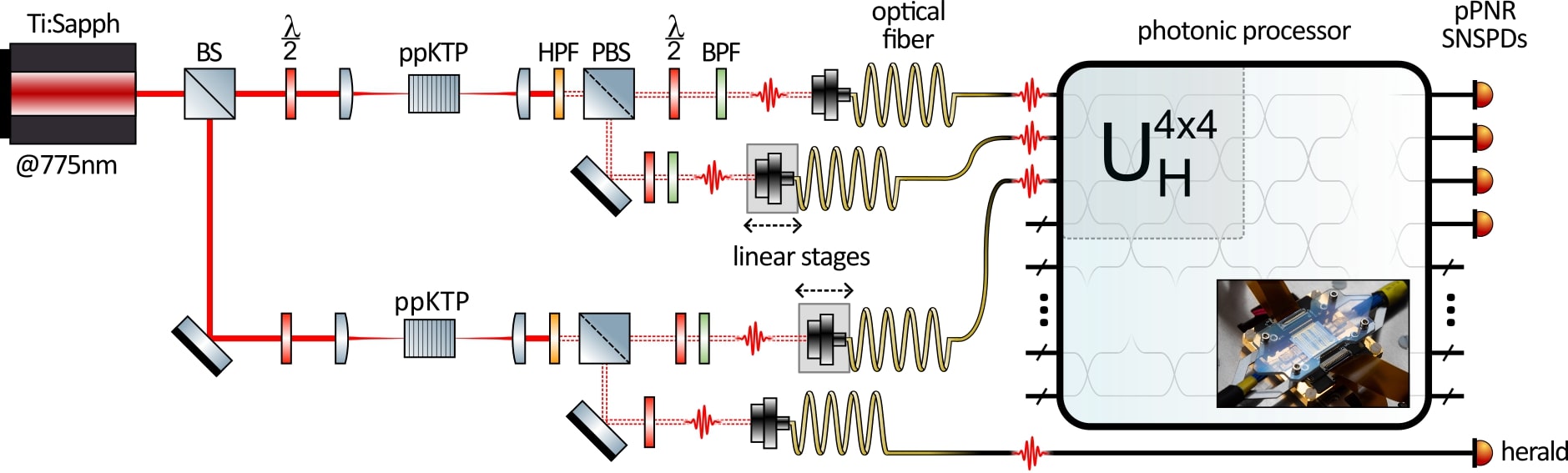}
    \caption{\textbf{Experimental Setup:} A Ti:Sapph laser pumps two single-photon sources in parallel. The beam is focused onto a ppKTP crystal to produce pairs of single photons (red wave-packets) via type-II SPDC, then filtered with a high-pass filter (HPF). The two photons are split at a polarizing beam splitter (PBS), filtered with bandpass filters (BPF), and coupled into optical fibers. Linear stages control the path lengths and thus the relative delays of the photons, to tune their partial distinguishability. The first three photons are sent to the photonic chip for the experiments, while the fourth is used as a herald. A 4-mode Haar-random unitary transformation ($U_H^{4\times 4}$) is programmed in the top-left corner of the chip, where the photons interfere. A picture of the photonic chip is included as an inset. The output state is detected using a bank of pseudo-photon-number-resolving superconducting nanowire single-photon detectors (pPNR SNSPDs).}
    \label{fig:Fig2_ExpSetup}
\end{figure*}

\subsection{Signatures of indistinguishability in Haar-averaged distributions}
\label{sec:theory-HaarAverage}

While so far we have focused on the validation of particular boson sampling experiments, where it is assumed that the unitary characterizing the interferometer is known \emph{a priori}, it is natural to ask whether signatures of multiphoton interference are also present in the Haar-averaged binned distributions. 

This question was tackled by \valery in \cite{shchesnovich2017quantum, shchesnovich2017asymptotic} in the two extreme scenarios of fully distinguishable vs. fully indistinguishable bosons, for an asymptotically large number of particles. Under generic assumptions, the probability of finding a configuration $\vect{k}$ is asymptotically given by a multivariate Gaussian distribution, with a larger standard deviation in the case of indistinguishable bosons. 
Despite being of fundamental interest, this result is of limited use for comparison with small-scale experiments boson sampling experiments. 
We present a new expression for the Haar-averaged variance of the photon-number distribution observed in a single bin $\mathcal{K}$
\begin{equation}
    \sigma_{\parti}^{2}(S)=\sum_{k=0}^{n}k^{2}\cdot P(k)-\left(\sum_{k=0}^{n}k\cdot P(k)\right)^{2}.
\end{equation} 
When averaged over the Haar measure for interferometers of $m$ modes, we obtain 
\begin{align}\label{eq: VarHaar}
    \left<\sigma_{\parti}^{2}(\mathcal{X})\right>_U=\frac{|\parti|n(m-|\parti|)(m^{2}-n)}{m^{2}(m^{2}-1)}+ \nonumber \\
    +\frac{m|\parti|-|\parti|^{2}}{m(m^{2}-1)}\sum_{i\neq j}|x_{ij}|^{2},
\end{align}
which naturally only depends on $n$, $m$, the size of the subset $|\setparti|$, and the sum over pairwise overlaps between internal wavefunctions of the input photons $|x_{ij}|^2= |\braket{\phi_i|\phi_j}|^2$. For detailed derivation we refer to Appendix \ref{sec: derivation of Var}. The expression is valid for any number of photons and confirms the behavior observed numerically in \cite{seron2022efficient}, showing that binned-mode photon number distributions become wider as the bosons become more and more indistinguishable.  This makes the variance of binned-mode distributions a good indicator of bosonic indistinguishability, even for small-scale boson sampling experiments. This behavior is also verified experimentally in Sec.~\ref{sec:res-phys}.

\section{Experimental setup}
\label{sec:exp}

The experimental setup outlined in Fig.~\ref{fig:Fig2_ExpSetup} is designed to enable high-precision experiments into multiphoton quantum interference, specifically in the context of boson sampling experiments targeted by the validation techniques introduced in Sec.~\ref{sec:theory}. It comprises three main components. The first is a photon source based on periodically poled potassium titanyl phosphate (ppKTP) crystals which generates high-purity single photons at telecom wavelengths, with the setup allowing for tunable degrees of partial distinguishability. These photons are then injected into a large-scale linear optical interferometer, implemented within an integrated photonic processor using silicon nitride waveguides. Finally, the interferometer's outputs are detected by a bank of superconducting nanowire single-photon detectors (SNSPDs), arranged to allow pseudo-photon number resolution for multiphoton events.
\\

Our photon source consists of a pair of periodically poled potassium titanyl phosphate (ppKTP) crystals configured in a Type-II degenerate setup, down-converting light from pulses of a titanium-sapphire (Ti:Sapph) pump laser centered at $775$nm to pairs of single photons at $1550$nm \cite{evans_2010_Phys.Rev.Lett.}, with an output bandwidth of approximately $\Delta \lambda = 20$nm. To enhance the purity of the two-photon state, the photons are filtered using bandpass filters (BPF) with a bandwidth of $\Delta \lambda = 12$nm. We use a single external herald detector and, by conditioning on the detection of three photons after the chip, we post-select on observing outcome $|\rangle$ \cite{tillmann_2013}, where the internal degrees of freedom are omitted in the notation for clarity. The photons are then coupled into optical fibers and directed to the photonic processor. By adjusting the relative arrival times of the photons using linear stages on the fiber couplers, we can continuously control their degree of distinguishability. On-chip measurements using the Hong-Ou-Mandel (HOM) effect \cite{hong_measurement_1987} provide a set of calibration measurements to infer the wave function overlap between photons $x_{ij} = \braket{\phi_i | \phi_j}$, where $\ket{\phi_i}$ represents the wave function of photon in  input spatial-mode $i$, as a function of the linear stage offset configuration (see Appendix \ref{sec:app-exp_partDist_tunig}). The maximum wave function overlap between photons, related to the HOM dip visibility via $V = x^2$ was measured. After filtering, we measure visibilities of 98\%, 95\% and 90\% for photons pairs 1\&2, 1\&3 and 2\&3, respectively (number refers to input spatial mode). It is important to note that HOM tests only provide access to $|x_{ij}|^2$. Therefore, we make the additional assumption that $x_{ij}$ is real. While this assumption may not hold in general, it applies to cases where partial distinguishability is introduced by time-delays, as verified in \cite{menssen2017distinguishability,rodari2024semideviceindependentcharacterizationmultiphoton}.

Our photonic processor consists of an interferometer implemented using silicon nitride waveguides \cite{TriPlex, QuiX2021}, with a total of $n = 12$ modes and an optical insertion loss (coupling plus propagation losses) of $\approx 5$dB ($68\%$), on average for the various input channels. Reconfigurability is achieved through an arrangement of unit cells, each consisting of pairwise mode interactions realized as tunable Mach-Zehnder interferometers \cite{clements_optimal_2016}. Each unit cell is adjusted via the thermo-optic effect. For a complete 12-mode transformation, the average amplitude fidelity is $F = {n}^{-1} \mathrm{Tr} (|U^{\dagger}_{\rm set}||U_{\rm get}|) = 90.4\pm 2.4\%$, where $U_{\rm set}$ and $U_{\rm get}$ are the intended and achieved unitary transformations, respectively. The processor also preserves the second-order coherence of the photons \cite{QuiX2021}. While the processor in principle allows for transformations between 12 modes, only a subsection of size 4 modes is used in these experiments due to limitations in the number of detectors available.

Photon detection is accomplished using a bank of 13 superconducting single-photon detectors \cite{ReviewSNSPD, Marsili-SNSPD}, with readout via standard correlation electronics. For each of the four modes of interest, three detectors are multiplexed to achieve pseudo-photon number resolution (pPNR) \cite{feito_measuring_2009}, with the thirteenth detector serving as the herald. We postselect on triple detection events, and correct our count rates for the reduction in effective efficiency due to the use of pPNR detectors \cite{somhorst_quantum_2023}. Due to this postselection, an additional source of noise arises: the generation of an extra photon pair followed by two optical loss events. The chosen photon generation probability represents a compromise between maximizing the overall generation rate and minimizing the impact of this noise, which we estimate to contribute less than 1\% to the relative error and therefore consider insignificant. 
\FloatBarrier

\begin{figure*}[t]
    \centering
        \includegraphics[width=0.85\textwidth]{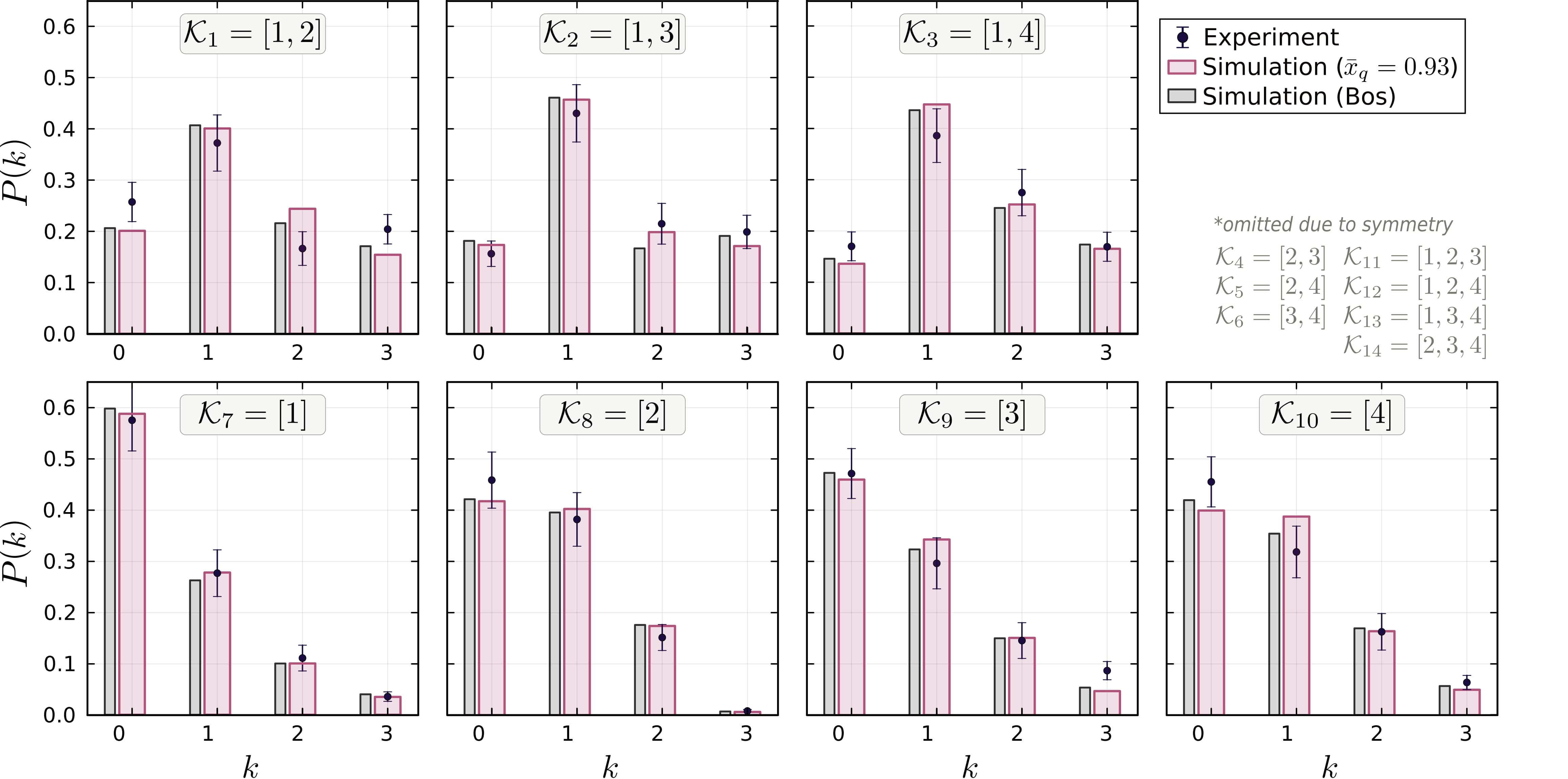}
    
    \caption{Binned-mode photon-number probability distributions for a single Haar-random unitary $U_H^1$, analyzed across 7 different mode binnings $\{\mathcal{K}_i\}$ with $i \in \{1, 2, 3, 7, 8, 9, 10\}$. Black circles represent experimental data with minimal partial distinguishability ($\bar{x}_q \approx 0.93$), and error bars indicate statistical noise. Pink bars correspond to numerical simulations with identical partial distinguishability to the experiment ($\bar{x}_q = 0.93$), while gray bars, shown to their left, depict simulations under perfectly bosonic conditions ($\bar{x}_q = 1$). Symmetrically equivalent binnings are omitted for clarity.}
    \label{fig:Fig3_singleU_Pk}
\end{figure*}

\section{Experimental validation}
\label{sec:res}

In this section, we present the results of the boson sampling experiments conducted on our integrated photonic platform. These experiments involved estimating the photon number probability distributions at the output modes of the chip through direct sampling. We performed measurements for 50 $4 \times 4$ random unitary transformations drawn from the Haar measure, denoted by the set of unitary matrices $\{U_H^i\}$, where $i \in {1, \dots, 50}$. For each transformation, the photon distinguishability was systematically varied at 9 different settings, ranging from as indistinguishable as possible in our setup to perfectly distinguishable photons, resulting in a total of 450 experiments.

Each unitary transformation was sampled for approximately 15 minutes at an average rate of 1 Hz, yielding an estimation error (standard deviation of the mean) of approximately 10\% per click pattern on average. The degree of distinguishability between photons for a given delay configuration was characterized by the set of pairwise overlaps $x_{ij} = \braket{\phi_i | \phi_j}$, from which the corresponding Gram matrix $\mathcal{X}$ was constructed. To encapsulate the degree of distinguishability, which depends on the three pairwise overlaps, into a single representative metric, we selected the quadratic mean, defined as
\begin{equation}\label{eq:quad_mean}
    \Bar{x}_q = \sqrt{x_{12}^2 + x_{13}^2 + x_{23}^2}.     
\end{equation}
This metric provides an effective representation of the overall distinguishability and is sensitive to inhomogeneities between photon pairs, making it well-suited for analyzing these experiments within our setup.

To validate our experimental results, we estimate the binned-mode distributions from the experimentally measured photon number probability distributions and compare with their simulated counterparts (see Sec. \ref{sec:theory}), which are generated using the Julia package \textsc{BosonSampling} \cite{seron2022bosonsampling}. Simulations employed the same experimental configurations, particularly, 
the input of $n = 3$ photons into the first $n$ modes and the set of unitary matrices $\{U_H^i\}$ and used exact methods to compute the binned-mode distributions. The partial distinguishability was varied accordingly, depending on the specific plot or comparison of interest, leveraging the full control over the pairwise overlaps between photons in the simulation.

\subsection{Binned-modes validation test }
\label{sec:res-val}

\emph{Single-unitary validation - }Figure~\ref{fig:Fig3_singleU_Pk} shows the binned-mode probability distributions $P(\mathbf{k})$ for three photons interfering in the 4-mode interferometer, configured with a single Haar-random unitary transformation. For this analysis, we selected $U_H^1$ from the set $\lbrace U_H\rbrace$ of 50 sampled unitaries. The dataset corresponds to the highest degree of indistinguishability achieved experimentally, with $\Bar{x}_q \sim 0.933$. Black circles indicate the experimentally estimated probabilities $P(\mathbf{k})$ of detecting $\mathbf{k}$ photons in the subset $\mathcal{K}_j$, with error bars reflecting statistical noise. Pink bars represent numerical simulations under the same level of distinguishability as the experiment ($\Bar{x}_q = 0.933$), while gray bars show numerical simulations for the ideal bosonic case ($\Bar{x}_q = 1$). The distributions are shown for various binning configurations of the interferometer's output modes. Given the small number of modes, we do not consider joint distributions over multiple bins. Instead, we measure the probability of detecting $k$ photons in a given bin, which implies $n-k$ photons in the complementary bin due to postselection on $n$-photon events. It is worth noting that certain binnings are omitted due to symmetry considerations: for example, the distribution for $\mathcal{K}_7 = [1]$ is symmetric to that of $\mathcal{K}_{14} = [2,3,4]$, and similar symmetries apply to other binnings.

This validation technique demonstrates its effectiveness even at the single-unitary level, where only one Haar-random transformation is analyzed. The binned-mode probability distributions $P(\mathbf{k})$ of an ideal boson sampler exhibit substantial differences between different binning configurations \cite{seron2022efficient, Singh2025_Pow_BS}. Consequently, from an adversarial perspective, as arises in proof-of-work applications \cite{Singh2025_Pow_BS}, it is challenging to spoof this validation test, as the choice of the binning can be done only after the data is received and the number of possible binning configurations grows exponentially with the size of the system. 

The experimental data reproduces well the  fluctuations between different binning choices, expected to happen in an ideal boson sampler. These fluctuations can be quantified by looking at TVD over different binning choices, 
\begin{equation}
   TVD(\mathcal{K}_i, \mathcal{K}_j)= \frac{1}{2}\sum_\mathbf{k} | P^{th}_{\mathcal{K}_j}(\mathbf{k})-P^{th}_{\mathcal{K}_i}(\mathbf{k})|,
\end{equation}   
for subsets $\mathcal{K}_i$ and $\mathcal{K}_j$ of equal size. For our experiment of three photons in four modes with unitary $U^1_H$, average value of $TVD(\mathcal{K}_i, \mathcal{K}_j)$ over all possible choices of two-mode bins is approximately 0.148. A naïve cheating strategy which would always generate the same binned probability distribution can thus be easily identified. The experimental data does significantly better than this naïve cheating strategy, since the average value of   
\begin{equation}
    TVD^{exp}_{\mathcal{K}_i} = \frac{1}{2}\sum_\mathbf{k} | P^{exp}_{\mathcal{K}_i}(\mathbf{k},\bar{x}_q)-P^{th}_{\mathcal{K}_i}(\mathbf{k})|
\end{equation}
over two-mode binnings is approximately 0.065. 

It is also natural to ask whether the experiment performs better than the uniform sampler with respect to this validation test. The uniform sampler would reproduce perfectly the Haar-averaged behavior of any binned-mode probability distribution, since the Haar-averaged value of any outcome probability is the same \cite{aaronson2011computational}. However, the uniform sampler fails to reproduce the dependency of the binned distributions on the specific unitary implemented, as well as on the specific bin choice. Focusing again on unitary $U_H^1$ and averaging over the 6 possible choices of two-mode bins, the expected value of the TVD between the Haar-averaged binned-mode distributions and the respective theoretically predicted distributions for unitary $U_H^1$ is approximately 0.191. This value is significantly larger than the previously reported value of 0.065, achieved by the experimental data.

The choice of the matrix $U_H^1$ was made arbitrarily, and a similar analysis can be done for the other 49 interferometers, drawn from the Haar ensemble, that were experimentally implemented. In Appendix \ref{sec:app_TVDs50Haar}, we report on the distribution of the measured values of the TVDs between the experimental and simulated binned-mode probability distributions for all the 50 interferometers and different bin choices. The results generally show a good agreement between the experimental observed behavior and that of an ideal boson sampler, with TVD values ranging between a minimum of 0.003 and a maximum of 0.195.

The analysis presented here serves as an important first step towards the validation of boson sampling in more complex schemes, such as those proposed for quantum proof-of-work applications \cite{Singh2025_Pow_BS}. A necessary, though not sufficient, condition for passing these validation tests is the ability to accurately reproduce the dependence of the binned-mode probability distributions on both the unitary transformation and the choice of binning. The experimental results demonstrate that our boson sampler successfully captures these fluctuations, distinguishing it from naïve cheating strategies, such as fixed-output distributions, and from uniform sampling, which fail to account for unitary-dependent correlations in the output. These findings pave the way for further investigations into more refined validation techniques, particularly those aimed at distinguishing experimental boson samplers from increasingly sophisticated classical mock-up samplers.

\begin{figure}[t]
    \centering
    \includegraphics[width=0.48\textwidth]{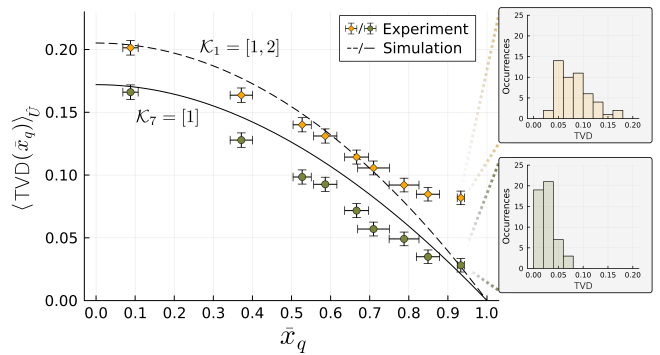}
    \caption{ Total variation distance (TVD) between probability distributions arising from partially distinguishable and fully indistinguishable photons on the binnings $\setparti_1 = [1,2]$ and $\setparti_7 = [1]$, averaged over 50 Haar random matrices, as a function of the partial distinguishability, measured by the quadratic mean of the photon pair wave-function overlaps $\Bar{x}_q$. Circles represent experimental data, with x-axis error coming from HOM visibility uncertainty and y-axis error given by error propagation. The black (solid and dashed) curves is the theoretical curve (zero imperfections), obtained simulating $10^4$ Gram matrices $\mathcal{X}(x)$, by changing the \textit{x} parameter continuously. Inset: histograms of TVD spread for the experiment with minimal partial distinguishability.}
    \label{fig:Fig6averageTVD}
\end{figure}

\emph{Sensitivity to partial distinguishability - } Figure \ref{fig:Fig6averageTVD} serves as experimental verification that the method is indeed sensitive to partial distinguishability, hence being able to distinguish an ideal boson sampler from a partial distinguishable one. Specifically, the circles in the plot represent the average TVD between the ideal bosonic binned-mode distribution and the experimentally measured one for a given partial distinguishability value $\Bar{x}_q$. The average is computed  over the set $\hat{U}$ of the 50 Haar-random unitaries that were implemented experimentally, according to:
\begin{equation}\label{eq.tvd-av}
    \langle TVD (\Bar{x}_q ) \rangle_{\hat{U}} = 
    \frac{1}{N_U} \sum_{\hat{U}} \frac{1}{2} \sum_\mathbf{k} |P^\text{exp}_{x=\Bar{x}_q}(\mathbf{k}) - P^\text{th}_{x=1}(\mathbf{k})|.
\end{equation}
 Moreover, the points shown correspond to two specific choices of binnings $\mathcal{K}_1 = [1,2]$ and $\mathcal{K}_7 = [1]$. Error bars on the x-axis indicate uncertainties in the HOM visibility measurements (see Appendix~\ref{sec:app-exp_partDist_tunig}), while those on the y-axis reflect the propagated errors from experimental noise. The inset histograms show the spread of the TVD values across the 50 Haar-random unitaries for the minimal partial distinguishability case implemented, corresponding to a value of $\Bar{x}_q \sim 0.933$. In turn, the solid and dashed curves are numerical simulations of the theoretical behavior  of the average TVD between binned-mode distributions of ideal and partially distinguishable particles, using the same bin choices and the same set of 50 Haar-random unitaries considered for the experimental points (in analogy to Eq.~\eqref{eq.tvd-av}). To obtain these curves we used a simple theoretical model of uniform partial distinguishability, defined by a single parameter $x$, corresponding to a Gram matrix $\mathcal{X}(x)$ such that $\mathcal{X}_{ij} = \langle \phi_i | \phi_j \rangle = x$ for all photon pairs $i, j$, and $\mathcal{X}_{ii}=1$.  By incrementing the $x$-parameter in steps of $10^{-4}$, the simulations continuously vary the state of the input photons from fully distinguishable particles ($x=0$) to completely indistinguishable ones $(x=1)$.

When examining the plot, we observe good agreement between the experimental data and the simulations at the lowest indistinguishability point ($\Bar{x}_q \sim 0$), where the experimental TVD closely matches the theoretical predictions. In the high-indistinguishability region (approximately the three rightmost points), the curves level out and saturate. This behavior indicates that in this regime, distinguishability ceases to be the dominant source of error. This effect is more pronounced for the bin of two modes $\mathcal{K}_1$, where the photon number distribution explicitly depends on the phases of the unitary transformation. In contrast, for the single-mode bin case $\mathcal{K}_7$, the distribution depends solely on the magnitudes $|U_{ij}|$. Further justification of this distinction is provided in Appendix~\ref{app:phase}. This suggests that the primary source of error arises from the imperfect control of the photonic processor during the programming of unitary transformations.

In the middle range of partial distinguishability, an interesting anomaly emerges. The TVD, as a metric, is expected to increase in the presence of any noise, meaning regions below the simulation curves should be inaccessible. However, the experimental data points fall below the theoretical curves in this region. We attribute this to an underestimation of the visibilities, which are determined through HOM dip measurements. This discrepancy is most pronounced in the middle range of partial distinguishability, where the HOM dip curves exhibit their steepest slopes, making visibility measurements particularly sensitive to drift and statistical noise (see Appendix \ref{sec:app-exp_partDist_tunig}).

This observation suggests an alternative approach for assessing distinguishability: directly comparing the experimental results to the simulated curves \cite{van2021experimental}. This method could provide a more robust lower bound on the partial distinguishability, avoiding the sensitivities and potential inaccuracies inherent in HOM dip measurements.
\\

\FloatBarrier

\emph{GBP measurements - }Additionally, we analyze the behavior of the \textit{generalized bunching probabilities} $P_3(\mathcal{K})$ defined in Eq. \eqref{eq:bunching_probability}, which quantify the likelihood of all photons bunching together in any given partition $\mathcal{K}$, comparing experimental results to numerical simulations.

\begin{figure}[t]
    \centering
        \includegraphics[width=0.48\textwidth]{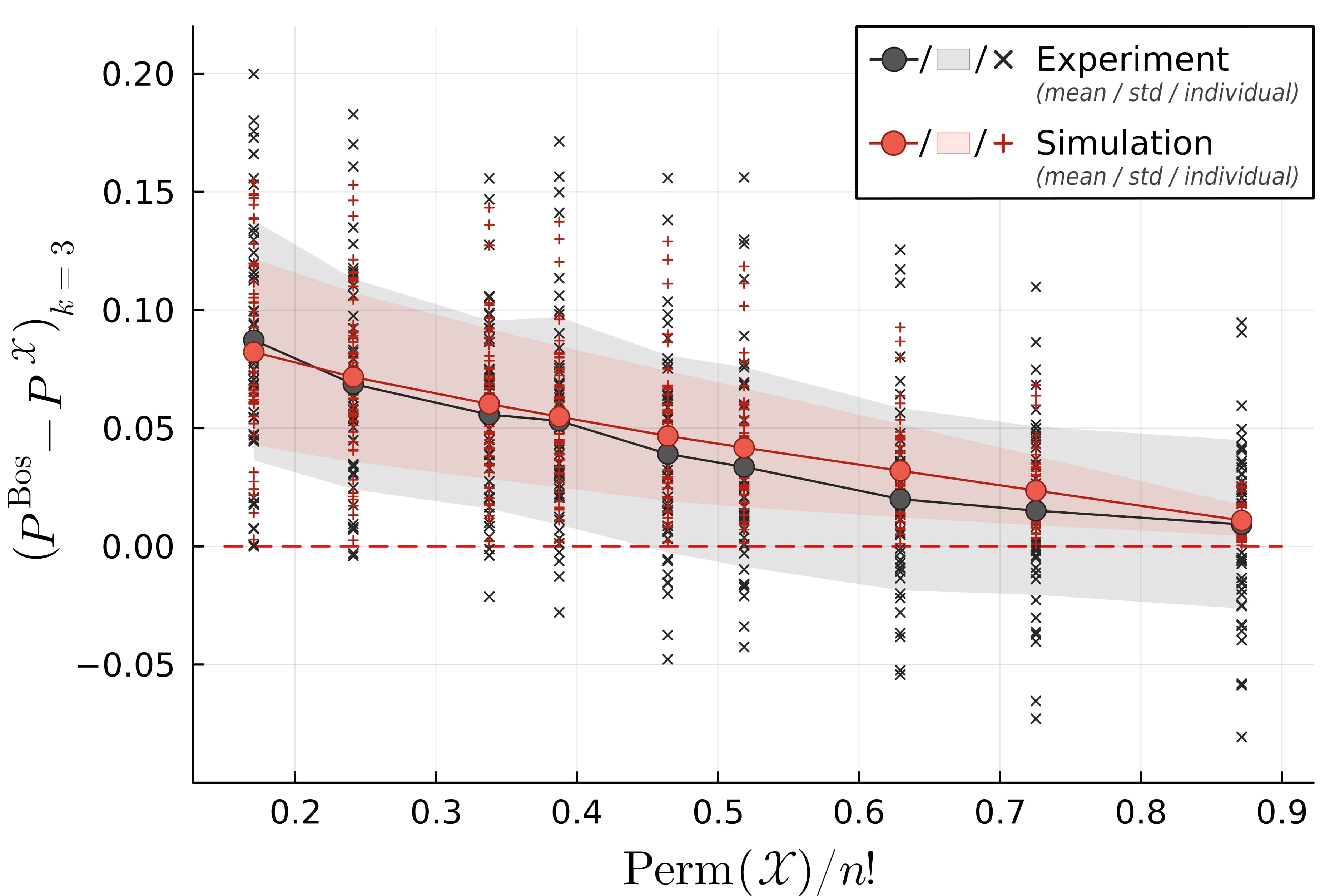}
    \caption{Difference between the generalized bunching probability (g.b.p.) at full indistinguishability ($P^\text{BOS}$, bosonic) and the g.b.p. at a given partial distinguishability ($P^\mathcal{X}$), plotted as a function of partial distinguishability ($\text{Perm}(\mathcal{X})/n!$), for all 50 random unitaries measured, for a fixed subset $\mathcal{K}_1 = [1,2]$. Black markers represent experimental data, while red markers correspond to numerical simulations. The solid cicles/lines and shaded regions show the mean and standard deviation, respectively, for each level of distinguishability. The $\times$/+ crosses represent the data for each unitary $U_H^i$.}
    \label{fig:Fig4genBunching}
\end{figure}

Figure~\ref{fig:Fig4genBunching} shows the difference between the "ideal" GBPs, $P^{\text{BOS}}_{k=3}$, obtained via numerical simulations with perfectly indistinguishable (bosonic) particles, and the GBP at a given partial distinguishability, $P^\mathcal{X}_{k=3}$, derived from experimental or simulated data. In this figure, we have chosen a fixed two-mode subset  $\mathcal{K}_1 = [1,2]$.  Moreover, we have chosen to plot the data as a function of the value of $\text{Perm}(\mathcal{X})/n!$, which can be seen as a quantifier of bosonic indistinguishability \cite{tichy2015_partial_distinguishability, shchesnovich2015tight}. The figure includes this difference for each of the 50 Haar-random unitaries in the set $\hat{U}$, alongside the mean and one standard deviation across $\hat{U}$.

This difference $(P^{\text{BOS}} - P^\mathcal{X})_{k=3}$ is expected to remain non-negative for three-photon experiments, for any choice of interferometer and any choice of subset (see Sec.~\ref{sec:theory-GBP}). While this behavior is consistently observed in the simulated data, the experimental data occasionally exhibit negative values, indicating more bunching at partial distinguishability than in the ideal case. Further analysis, detailed Appendix \ref{sec:app-NoisyMatrixGBP}, shows that this effect arises due to noise in programming the photonic chip, i.e., the dialed matrix on the device deviates slightly from the target unitary. Simulations incorporating this noise match the experimental data almost perfectly, providing strong evidence that the observed discrepancies stem from this source. Moreover, in Appendix~\ref{sec:app-NoisyMatrixGBP}, we show a similar plot for the differences between GBPs for a single-mode subset $\mathcal{K}_7=[1]$ (also referred to as full bunching probability \cite{general_rules_bunching, rodari_experimental_2024}). In this case, the value of these probabilities should not depend on the phases of matrix elements $U_{ij}$ and, indeed, much fewer negative values are experimentally observed. This further suggests that the mischaracterization of the phases of $U_{ij}$ is responsible for the deviations between the theory and experiment observed in Fig.~\ref{fig:Fig4genBunching}.

\subsection{Haar averaged behavior}
\label{sec:res-phys}

We now turn to the analysis of results averaged over the Haar ensemble. This approach provides a statistical perspective on multiphoton interference, revealing universal features that persist beyond the specifics of individual unitaries. By averaging over a large set of Haar-random unitaries, we can identify general trends while suppressing noise associated with unitary-specific behaviors. One such universal feature is the sensitivity of multiphoton interference to partial distinguishability, which emerges as a particularly visible and relevant aspect in our analysis. This sensitivity shapes the behavior of binned-mode probability distributions, underscoring its fundamental role in validating boson sampling experiments.
\\

\begin{figure*}[t]
    \centering
        \includegraphics[width=0.92\textwidth]{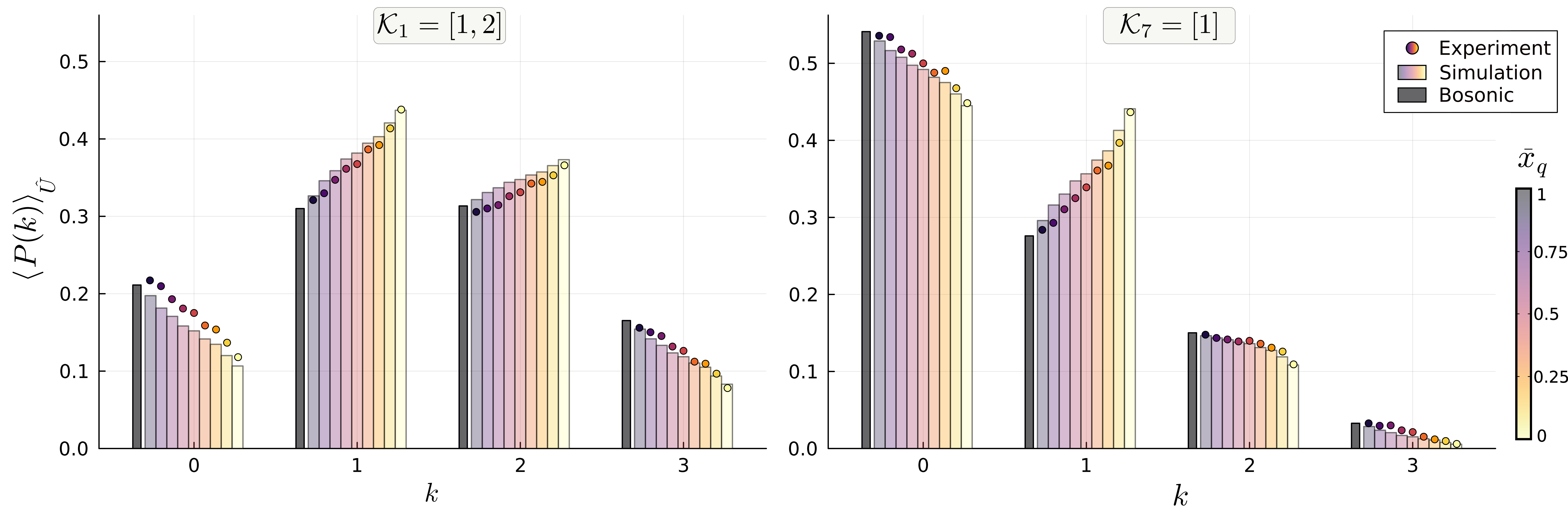}
    \caption{Averaged probability distribution on binnings $\setparti_1 = [1,2]$ and $\setparti_7 = [1]$, arising from partially distinguishable photons. The average is over 50 Haar-random unitaries. Circles represent the experimental data, ordered from left-to-right with increasing distinguishability, for each \textit{k}. The bars represent the corresponding simulated results, computed with the same sampled unitaries and Gram matrices. The partial distinguishability is represented by the quadratic mean of the photon pair wave-function overlaps $\Bar{x}_q$, indicated by the colorbar. Error bars too small to be visible.}
    \label{fig:Fig5averageDist}
\end{figure*}

Figure~\ref{fig:Fig5averageDist} shows the averaged photon-number probability distributions $\langle P(k) \rangle_{\hat{U}}$ for binnings $\mathcal{K}_1 = [1,2]$ and $\mathcal{K}_7 = [1]$, averaged over the set $\lbrace U_H \rbrace$ of 50 Haar-random unitaries. Circles correspond to experimental data, colored by the quadratic mean of the photon pair wavefunction overlaps, $\Bar{x}_q$, as indicated by the colormap on the right. Bars represent numerical simulations, with black bars corresponding to the ideal bosonic case ($\Bar{x}_q = 1$). Error bars representing one standard deviation of the mean are plotted but are too small to be visible. Due to the symmetry of the Haar ensemble, all partitions of identical size exhibit the same behavior after averaging, making it sufficient to study only the two selected partitions, $\mathcal{K}_1 = [1,2]$ and $\mathcal{K}_7 = [1]$, to fully encapsulate the effects of partial distinguishability on Haar-averaged distributions. 

These results provide a clear example of how partial distinguishability affects the binned-mode photon number distributions. Specifically, we observe a clear trend where the Haar-averaged frequency of full bunching events ($k = 3$) decreases monotonically as distinguishability increases. As a consequence of photon-number conservation, this also implies that the probability of observing vacuum - $k = 0$ - decreases, since it is equal to the probability of bunching in the complementary subset. This general tendency of boson to bunch results in a gradually widening of the distribution throughout this classical-to-quantum transition, governed by the tunable parameter $\Bar{x}_q$.  


It is important to remark that this trend does not necessarily arise when examining these distributions for single unitaries rather than ensemble averages. In some cases, one observes non-monotonic behavior, where the probability of detecting $k$ photons in a given partition does not peak at maximum or minimum distinguishability but rather somewhere in between. While this phenomenon is not particularly relevant for validation purposes, it is an interesting feature of multiphoton interference and relates to a discussion initiated by Tichy et al. \cite{tichy2011four}. Further details and examples of this behavior can be found  the Appendix \ref{sec:app-p(k)}. 
\\

\begin{figure}[t]
\flushleft
    \includegraphics[width=0.45\textwidth]{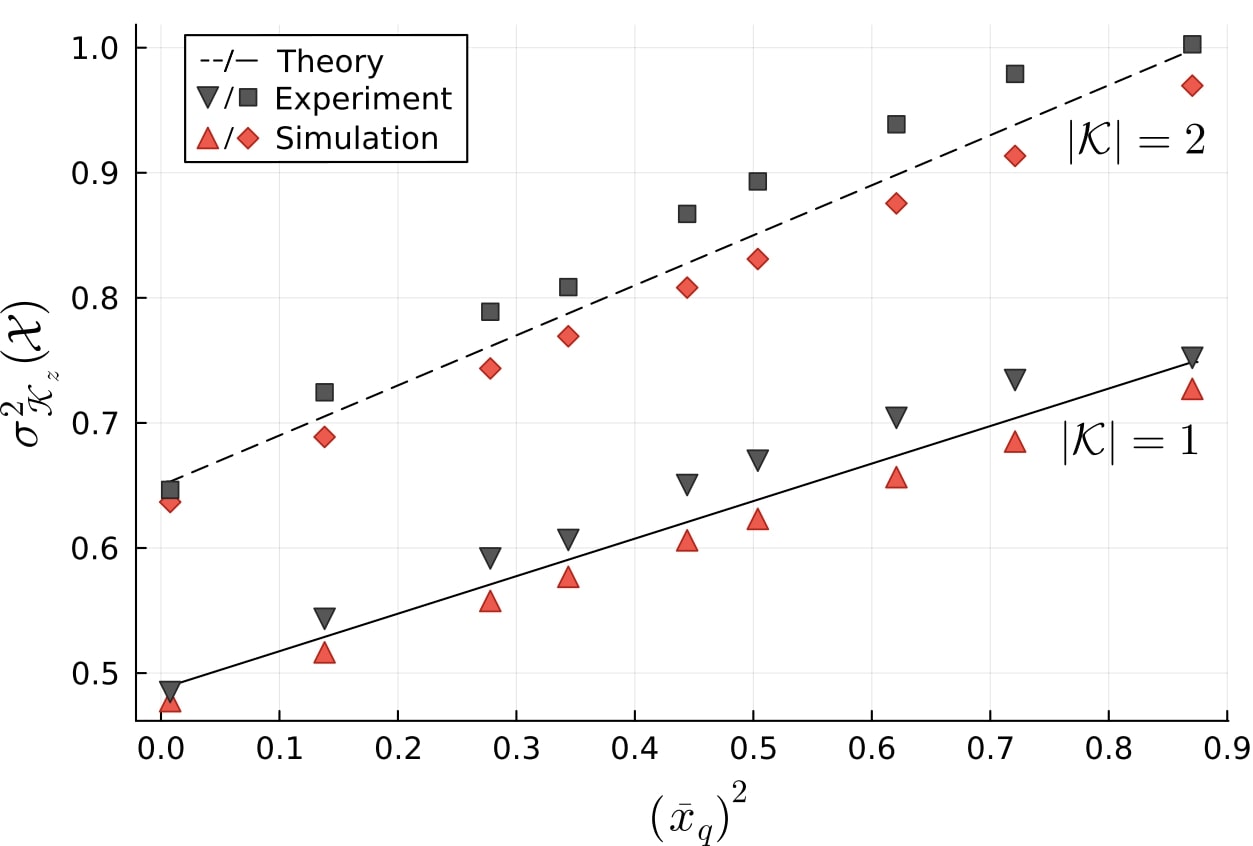}
    \caption{Variance of the photon-number probability distributions $\sigma^2_{\mathcal{K}_z}$, for two binning sizes as a function of partial distinguishability $(\Bar{x}_q)^2$. Solid and dashed lines represent the theoretical predictions, while black (inverted triangles/squares) and red (triangles/diamonds) markers correspond to experimental and numerical simulation data, respectively, for $|\mathcal{K}|=1$ and $|\mathcal{K}|=2$.}
    \label{fig:Fig7varInt}
\end{figure}

Figure~\ref{fig:Fig7varInt} shows the variance of the photon-number probability distributions, $\sigma_{\mathcal{K}_x}^2$ over the Haar ensemble (set $\lbrace U_i\rbrace$), as a function of the squared partial distinguishability parameter, $\Bar{x}_q^2$, for two binning sizes: $|\mathcal{K}| = 1$ and $|\mathcal{K}| = 2$. Solid and dashed lines represent the theoretical predictions for these two binning sizes, computed from Eq. \ref{eq: VarHaar}, while black markers (inverted triangles and squares) correspond to experimental data, and red markers (triangles and diamonds) represent numerical simulations. The variance is computed for both the experimental and simulated photon-number distributions, and the results are compared to the theoretical curves.

The variance of the photon-number probability distributions, as shown in Fig. \ref{fig:Fig7varInt}, serves as a strong indicator of photon indistinguishability, which can be used to directly extract the average HOM visibility between the photon pairs, given by the parameter $\Bar{x}_q^2$ . The results accurately confirm the theoretical prediction, which depends only on the number of binned modes $|\mathcal{K}|$, and is independent of the specific choice of the partition. The simulation points (red markers) closely follow the theoretical curves but consistently lie slightly below them. This deviation is attributed to the finite size of the sampled Haar-random unitaries set, $|\lbrace U_i \rbrace| = 50$, which introduces minor statistical effects.

Interestingly, the experimental points (black markers) lie above the theoretical curves, a surprising observation. We believe this discrepancy can be attributed to an underestimation of the distinguishabilities, as previously discussed. Notably, the experimental points that deviate above the curves are concentrated in the middle range of distinguishability, where the HOM visibility measurements are most sensitive to drift and statistical noise. This region aligns with where the underestimation of partial distinguishabilities has the greatest impact, while the points at the extreme ends of distinguishability exhibit a better agreement with the simulations and theory.

Our results show that measuring the variance of binned-mode distributions provides a practical method to infer the average pairwise visibilities of the input photons through simple post-processing of data coming from reconfigurable boson sampling experiments with Haar-random interferometers. Recently, measurement of photon-number variances of the marginal state of a single output mode has been proposed as a way to obtain guarantees on photonic indistinguishability \cite{rodari2024semideviceindependentcharacterizationmultiphoton}. Our results show that this approach is equally valid in the broader context of binned-mode distributions. This generalization is particularly significant for the scalability of this method, as the variance of binned-mode distributions should be easier to measure experimentally for larger boson samplers, especially in the  the low density (collision-free) regime where $m\gg n^2$, which is needed to show classical hardness of the boson sampling problem. In such a regime, examining a single output mode becomes challenging because it will most often register no photons and, less frequently, a single photon, leading to a highly sparse distribution. By contrast, binned-mode distributions allow the use of larger bins, where the probabilities $P(k)$ are more evenly distributed and non-negligible for higher photon numbers, resulting in a more reliable and less noisy variance measurement.

\section{Conclusions}
\label{sec:conclusion}

In this work, we have experimentally tested a method for assessing the performance of boson sampling experiments by leveraging binned-mode photon-number distributions. Our results confirm that this approach provides a scalable and efficient means to analyze multiphoton interference experiments, as it is sensitive to the presence of experimental sources of noise, such as partial distinguishability or imperfect control of the processor.

First, we demonstrate this validation technique at the level of a single unitary transformation, by direct comparison between experimental and simulated probability distributions on binned outputs. We show that this approach provides a straightforward yet stringent tool for evaluating the correctness of boson sampling experiments, since a good experimental implementation should correctly reproduce the variations between the different possible binned-mode distributions expected of an ideal boson sampler. These variations depend both on the specific interferometer as well as the choice of bin and we show that the experiment reproduces the expected behavior better than certain simple mock-up strategies.

In addition, we tested the sensitivity of this validation technique to partial distinguishability, by measuring binned-mode distributions for a total of 50 unitary transformations while tuning partial distinguishability via controlled delays. The average total variation distance (TVD) was used to quantify the deviation of the experimental data from the sampling behavior of the ideal bosonic case. The deviations between the ideal distributions and the experiment became more and more significant as the photons became more and more distinguishable, experimentally showing that the method can differentiate an ideal boson sampler from one with partially distinguishable particles.

This sensitivity to partial distinguishability is, in part, due to the fact that a particular probability outcome of binned-mode distributions are generalized bunching probabilities (GBP), which were also measured in our experiments. Our data generally confirms that these probabilities decrease with distinguishability for any given unitary and different subset choices \cite{shchesnovich2016universality}. However, we observed abnormalities in the experimental data, which we attribute to noise when programming the unitary transformation on the processor.

Furthermore, we use the ensemble of the data collected over 50 Haar-random interferometers to approximate Haar-averaged binned-mode distributions. We experimentally explored the transition between the behavior of distinguishable bosons, governed by classical statistics, and that of indistinguishable bosons showing the binned-mode distributions become wider with particle indistinguishability. We also observed a very good agreement to the theoretical prediction that the Haar-averaged variance of the binned photon-number distributions on the average pairwise HOM visibilities of the input photons. 


In summary, the techniques applied provide a versatile framework not only to validate the functionality of boson samplers, but also to serve as diagnostic tools for identifying and mitigating experimental imperfections. Future work can build on these results to extend the validation framework to larger photonic systems and more intricate experimental setups.


\section*{Data and code availability}

The numerical simulations for this project were performed with \href{https://github.com/benoitseron/BosonSampling.jl}{\textsc{BosonSampling.jl}} \cite{seron2022bosonsampling}.\\

The experimental data and processing scripts can be found in the repository "Data underlying the publication "Experimental validation of boson sampling using detector binning"" at 4TU (DOI: 10.4121/1ad5f239-5c41-4870-890a-91a2fa1b6653).

\section*{Acknowledgments}

This publication is part of the Vidi project \textit{At the Quantum Edge} which is financed by the Dutch Research Council (NWO). This work received support from HTSM-KIC project \textit{Building Einstein's Dice}. A.C. acknowledges financial support from FCT - Fundação para a Ciência e a Tecnologia (Portugal) via PhD Grant SFRH/BD/151190/2021. B.S. acknowledges funding from the Fonds National de la Recherche Scientifique – FNRS, as well as the Georg H. Endress Foundation. M.R. acknowledge funding from the European Union’s Horizon 2020 research and innovation programme under Marie Sklodowska-Curie grant agreement No. 956071 (AppQInfo) and from the Fonds de la Recherche Scientifique – FNRS. M.R. is a FRIA grantee of the Fonds de la Recherche Scientifique – FNRS. L.N. acknowledges support from FCT via the Project No. CEECINST/00062/2018 and from the European Union's Horizon Europe research and innovation program under EPIQUE Project GA No. 101135288.  L.N. also acknowledges the financial support of the project with the reference n.º 2023.15565.PEX, funded by national funds through FCT, I.P https://doi.org/10.54499/2023.15565.PEX

\section*{Competing Interests}
{The authors declare no competing interests.}


\section*{Appendix}

\appendix

\section{Characteristic function of binned-mode distributions}\label{sec:app_charfunction_permanent}
The (virtual) interferometer $\op{V}(\vect{\eta})=\op{U}^{\dagger} e^{i \vect{\eta}\cdot \vect{\hat{N}_{\mathcal{K}}}} 
\op{U}$ appearing in the computation of the characteristic function from Eq.~\eqref{eq:characteristic_interferometer} is characterized by a $m\times m$ unitary matrix $V(\vect{\eta})$, constructed as 
\begin{align}\label{eq:defV}
V(\vect{\eta})&=U^{\dagger} \Lambda(\vect{\eta})U, 
\end{align} 
where $\Lambda(\vect{\eta})$ is a diagonal matrix given by a product of diagonal matrices 
\begin{equation}
\Lambda(\vect{\eta})= \prod_{z=1}^K D^{(z)}(\eta_z), 
\end{equation}
such that 
\begin{equation}
\label{eq:dab}
    D^{(z)}_{ab}(\eta_z)= \begin{cases}
    e^{i \eta_z},~\text{ if}~a=b~\text{and}~a\in\setparti_z,\\
    1, ~\text{ if}~a=b~\text{and}~a\notin\setparti_z,\\
    0, ~\text{ if}~a\neq b.
    \end{cases}
\end{equation}
In the case where the initial state is a set of partially distinguishable photons, with one photon in each of the first $n$ modes, given by
\begin{equation}
   \ket{\Psi_\mathrm{in}}= \prod_{j=1}^m \op{a}_{j, \phi_j}^\dagger\ket{0},
\end{equation}
the expression for the characteristic function takes the form \cite{seron2022efficient} 
\begin{equation}
\label{eq:xl_perm}
x(\vect{\eta})=\perm{\mathcal{X}\odot V_n(\vect{\eta})}, 
\end{equation}
where $\odot$ is the Hadamard (elementwise) product: $(A\odot B)_{ij} = A_{ij}B_{ij}$. Here, $V_n(\vect{\eta})$ is a submatrix of $V(\vect{\eta})$ obtained by selecting its first $n$ rows and columns.

\section{Variance of binned distributions}\label{sec: derivation of Var}
In this section we derive the formula for the variance of the distribution of the number of photons in a single bin $\mathcal{K}$. We first show how the Fourier transform of this probability distribution can be expressed and then compute its statistical moments. 
As proven in previous work \cite{seron2022efficient}, the Fourier transform of $P(k)$ takes the form
\begin{equation}\label{eq: GF}
    F(y)=1+\sum_{a=1}^{n}c_{a}(e^{iy}-1)^{a},
\end{equation}
where the coefficients $c_a$ are given by
\begin{equation}\label{eq: coefficient GF}
    c_{a}=\sum_{\omega\in Q_{a,n}}\perm{(H\odot \mathcal{X})[\omega]}.
\end{equation}
Here, $Q_{a,n}$ denotes the set of all strictly ordered subsets of $\omega\subset \{1,...,n\}$ of $a$ elements and $(H\odot \mathcal{X})[\omega]$ denoted the submatrix of $H\odot \mathcal{X}$ containing the rows and columns picked according to $\omega$. With a change of variables, we can obtain the moment generating function as
\begin{equation}
    M(y)= F(-iy)=1+\sum_{a=1}^{n}c_{a}(e^{y}-1)^{a},
\end{equation}
such that
\begin{equation}
    M(y)=\sum_{k}P(k)e^{ky}\implies \frac{\partial^{m}}{\partial y^{m}}M(y){\bigg |}_{y=0}=\sum_{k}k^{m}P(k)
\end{equation}
Now it can be seen that
\begin{equation}
    \frac{\partial^{m}}{\partial y^{m}}(e^{y}-1)^{a}{\bigg |}_{y=0}=0 \ \ \text{for }a > m.
\end{equation}
To compute the variance of the distribution $P(k)$, we therefore only need to expand $M(y)$ up to order two and compute the corresponding statistical moments. Let us first note that the terms appearing in the computation of $c_1$ are given by
\begin{equation}
    \perm{(H\odot \mathcal{X})[\{i\}]}=H_{i,i}=\sum_{k\in \parti}U_{k,i}^{*}U_{k,i} \ \ \forall i,
\end{equation}
whereas the terms appearing in the computation of $c_2$ are
\begin{equation}
    \perm{(H\odot \mathcal{X})[\{i,j\}]}=H_{i,i}H_{j,j}+|x_{i,j}|^{2}H_{i,j}H_{j,i} \ \ \forall i\neq j.
\end{equation}
The last step to obtain the Eq.~\eqref{eq: VarHaar} is to average over the Haar measure. This can be done using the Weingarten function \cite{Weingarten}. In particular, given that $U\in \mathbb{C}^{m\times m}$,
\begin{equation}
    \int dU U_{i,j}U_{k,l}^{*}=\frac{\delta_{i,k}\delta_{j,l}}{m}
\end{equation}
\begin{align}
   \int dU U_{i,j}U_{k,l}U_{i',j'}^{*}U_{k',l'}^{*}&= \nonumber
   \\&\frac{\delta_{i,i'}\delta_{j,j'}\delta_{k,k'}\delta_{l,l'}+\delta_{i,k'}\delta_{j,l'}\delta_{k,i'}\delta_{l,j'}}{m^{2}-1} \nonumber\\
   -&\frac{\delta_{i,i'}\delta_{j,l'}\delta_{k,k'}\delta_{l,j'}+\delta_{i,k'}\delta_{j,j'}\delta_{k,i'}\delta_{l,l'}}{m(m^{2}-1)}
\end{align}
Using these last results, we obtain
\begin{align}
    \left<\sigma_{\parti}^{2}(\mathcal{X})\right>_U=\frac{|\parti|n(m-|\parti|)(m^{2}-n)}{m^{2}(m^{2}-1)}+ \nonumber \\
    +\frac{m|\parti|-|\parti|^{2}}{m(m^{2}-1)}\sum_{i\neq j}|x_{ij}|^{2}.
\end{align}
We note that a dependency on the average pairwise HOM visibilites of the input photons also appears in the photon number variances of single-mode output states after an unbiased interferometry process \cite{robbio2024centrallimittheorempartially, rodari_experimental_2024}. 

\section{Partial distinguishability tuning}\label{sec:app-exp_partDist_tunig}

\begin{figure*}[t]
    \begin{centering}
        \includegraphics[width=0.95\textwidth]{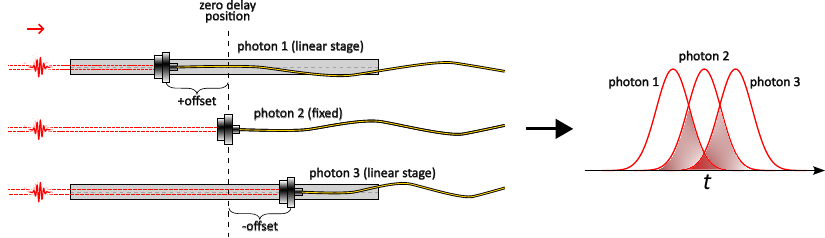}
        \caption{Schematic of the setup used to control photon delays by adjusting optical path lengths. Photon 1 serves as a fixed reference point with zero delay, while photons 1 and 3 are positioned on linear stages to allow for precise path length adjustments. Photon 1 is delayed by offsetting its path length in one direction, while photon 3 is offset by the same amount in the opposite direction. This configuration ensures comparable partial distinguishability for pairs 1\&2 and 2\&3, while pair 1\&3, subject to twice the offset, is significantly more distinguishable, as shown by the temporal overlap on the right side of the figure.}
        \label{fig:photonDelays}
    \end{centering}
\end{figure*}

To tune the partial distinguishability between our photons, we use the temporal degree of photons. To control the temporal overlap between photons, we adjust their delays by modifying the optical path lengths, as illustrated in Fig.~\ref{fig:photonDelays}. Photon 2 serves as a fixed reference with an unchanged path length, establishing the "zero delay" baseline. Photons 1 and 3, placed on SmarAct SLC-2490 linear stages, have adjustable path lengths, allowing precise delay control relative to photon 2. Photon 1’s delay is adjusted directly, while photon 3 is offset in the opposite direction, enabling control over the relative delays between photon pairs within the constraints of the system. This choice of delays ensures that while the partial distinguishability between photon pairs 1\&2 and 2\&3 is of a similar order, the delay offset for pair 1\&3 is twice as large, making it significantly more distinguishable. For this reason, we use the quadratic mean as a metric to encapsulate the average partial distinguishability across the three photon pairs, as it is more sensitive to larger deviations from the mean.

The coherence length of the photons, measured in a Hong-Ou-Mandel (HOM) experiment, is approximately 300 µm, while the SmarAct SLC-2490 stages have a resolution of about 10 nm, giving a relative "resolution to total length" control error of ~0.003\%. This high precision effectively provides arbitrary control over the temporal distinguishability between photon pairs. This setup, enabling precise control over relative delays, is essential for exploring the effects of partial photon distinguishability and HOM visibility on the resulting interference patterns.

\begin{figure*}[t]
    \begin{centering}
        \includegraphics[width=0.95\textwidth]{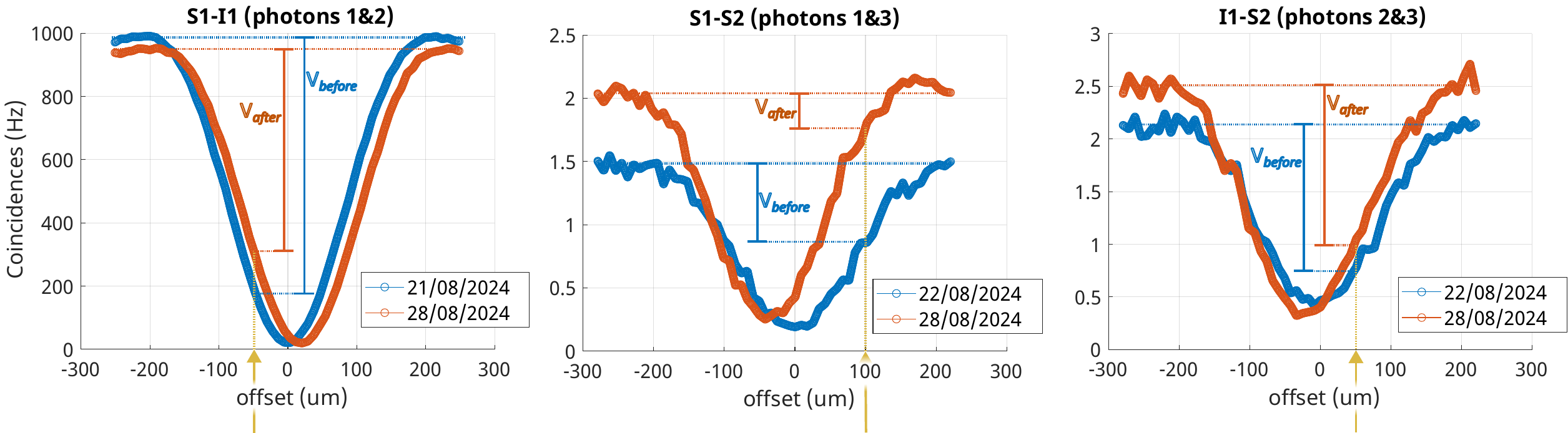}
        \caption{Experimentally measured HOM dips between all three photon pairs before (blue) and after (orange) the boson sampling validation measurements. The x-axis represents the offset in micrometers (µm), and the y-axis shows the coincidence count rate in Hz. From left to right, the plots show HOM dips for the photon pairs Signal1-Idler1 (same source), Signal1-Signal2, and Idler1-Signal2 (across sources). The figure highlights an example of the measured visibilities at a 50 µm offset configuration (indicated by the gold arrow and dashed lines), which provides information to infer the partial distinguishability.}
        \label{fig:HOMdips}
    \end{centering}
\end{figure*}

The primary sources of error in estimating partial distinguishability are statistical noise in the measured HOM dip curves and drift over time. The statistical noise is primarily due to low coincidence counts, around 1 Hz, between photons from different sources, as both photons must be heralded. Additionally, gradual drifts in the alignment of photon sources, the pump laser, and the optical path lengths introduce variability. These drifts occurred over the course of the measurement period, which ran from 22/08/2024 to 28/08/2024, with HOM dips measured on 21/08/2024 and 29/08/2024 to track changes (see Fig. \ref{fig:HOMdips}). To estimate partial distinguishability, we measure the full HOM dip curve, which contains information about all partial distinguishabilities as a function of offsets. Given the high accuracy of these offsets, we then retroactively infer partial distinguishability based on the observed visibility across the curve. Under the assumption that these drifts occur in a linear fashion on average, we infer that the true partial distinguishability at any given offset lies between the two extreme values observed in the HOM dips measured before and after the experiment. This drift represents the dominant systematic error; therefore, all x-axis errors in the quadratic mean values presented in this paper are calculated by taking the midpoint of these two partial distinguishability extremes, with the error bars determined by the range between the extreme values.

\section{Validation of 50 Haar-random interferometers}\label{sec:app_TVDs50Haar}
In this section we present complementary data regarding the validation of the experimental implementation of boson sampling with 50 different Haar random unitaries.  We focus on the implementation with the highest indistinguishability achieved in the experiment, corresponding to a parameter $\bar{x}_q\sim 0.993$ (see Eq.~\eqref{eq:quad_mean}), and compare the experimentally measured binned-mode distributions to the corresponding ones predicted for an ideal boson sampler. 

\begin{figure*}[t]
     \centering
    \subfigure[]{
        \includegraphics[width=0.32\textwidth]{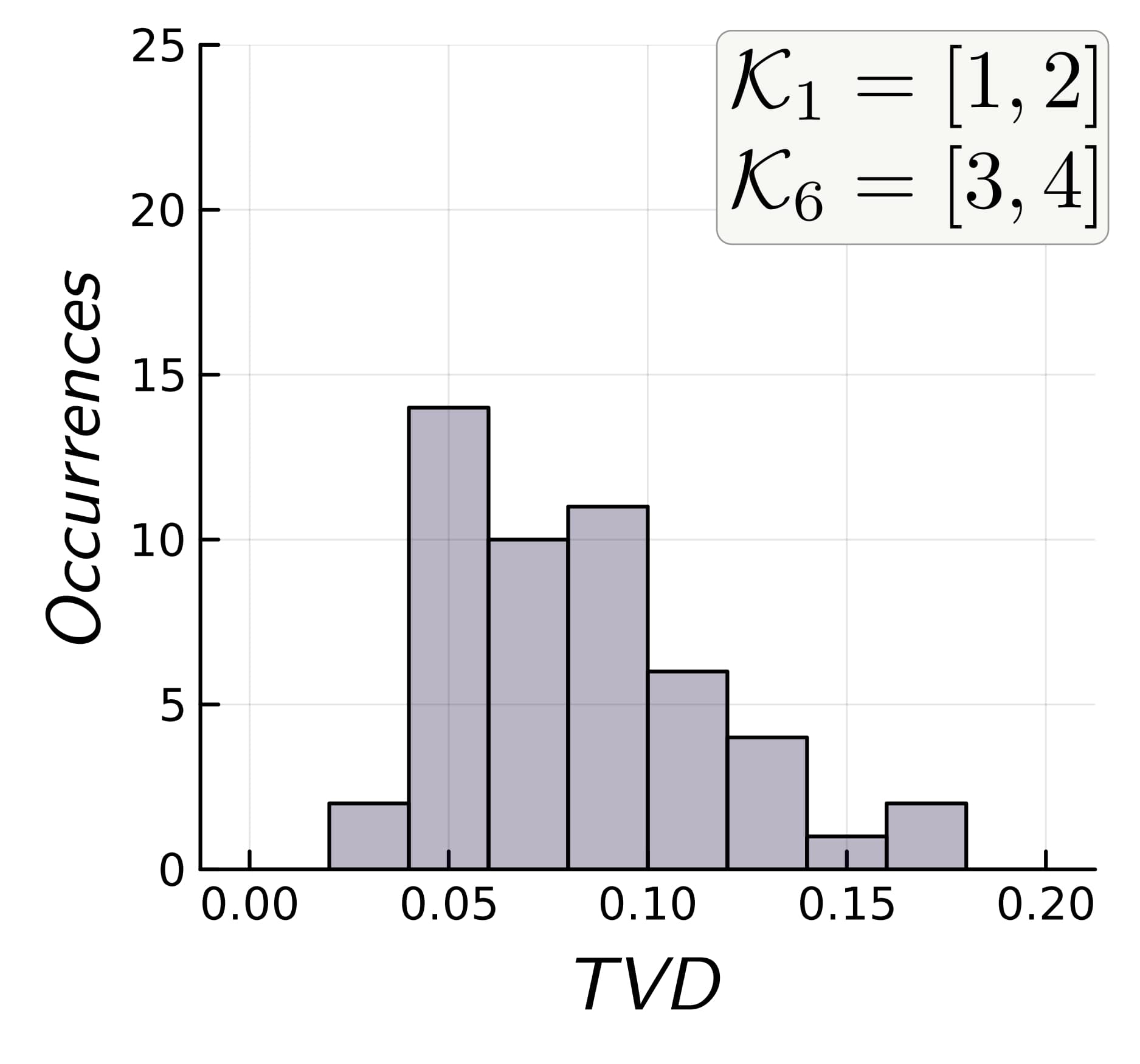}
    }
    \hfill
    \subfigure[]{
        \includegraphics[width=0.32\textwidth]{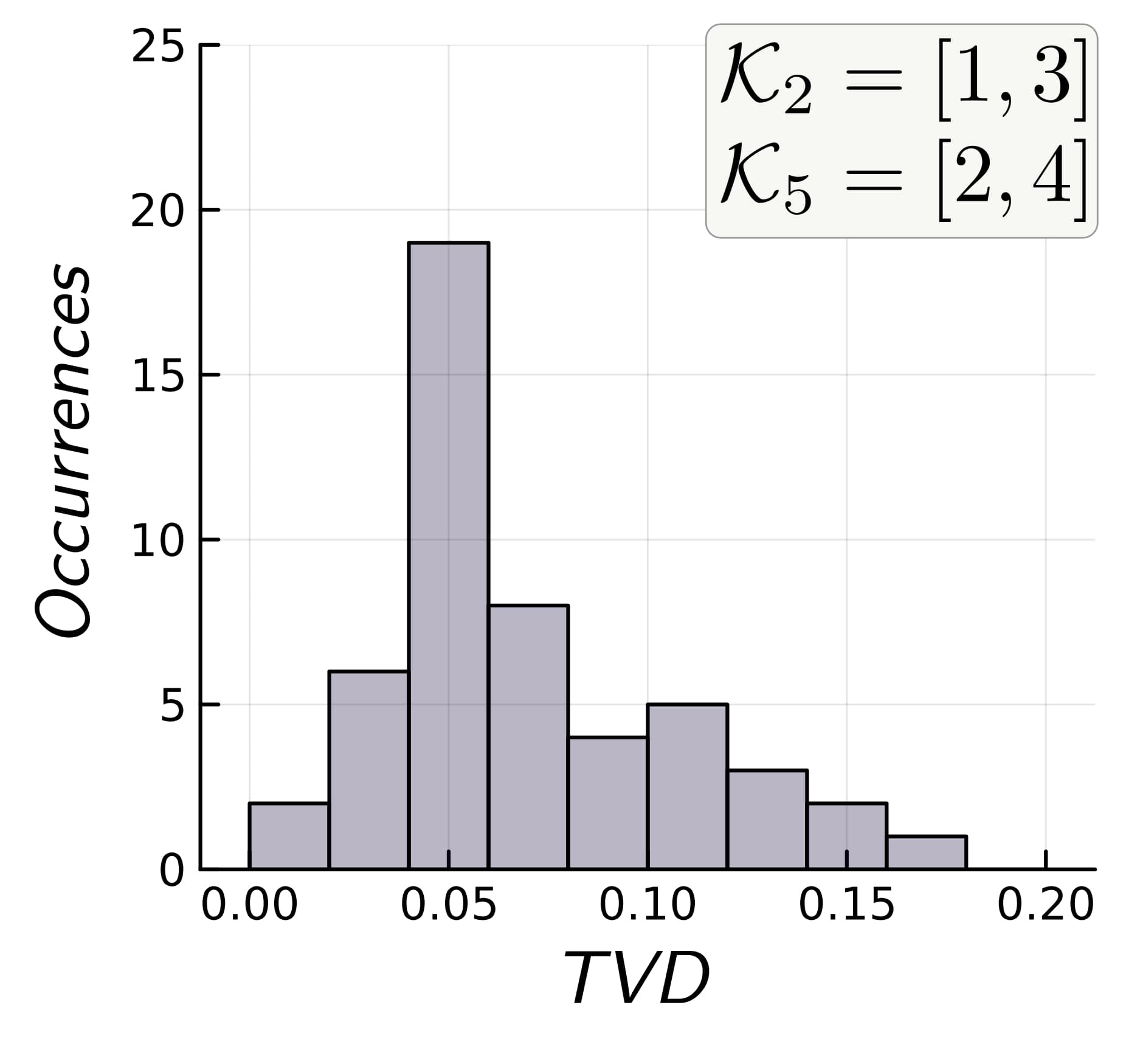}
    }
    \hfill
        \subfigure[]{
        \includegraphics[width=0.32\textwidth]{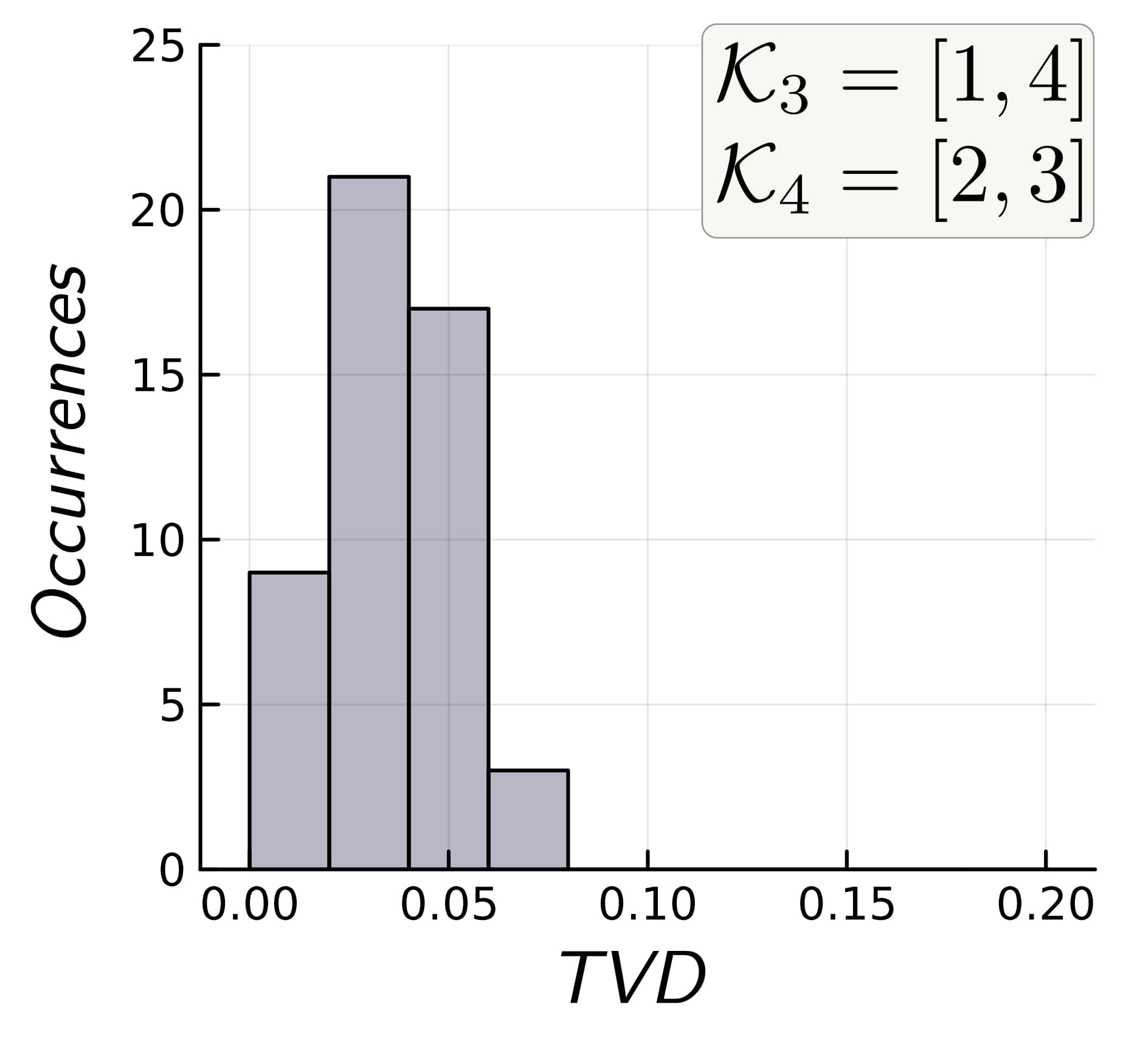}
    }
    \hfill
        \subfigure[]{
        \includegraphics[width=0.32\textwidth]{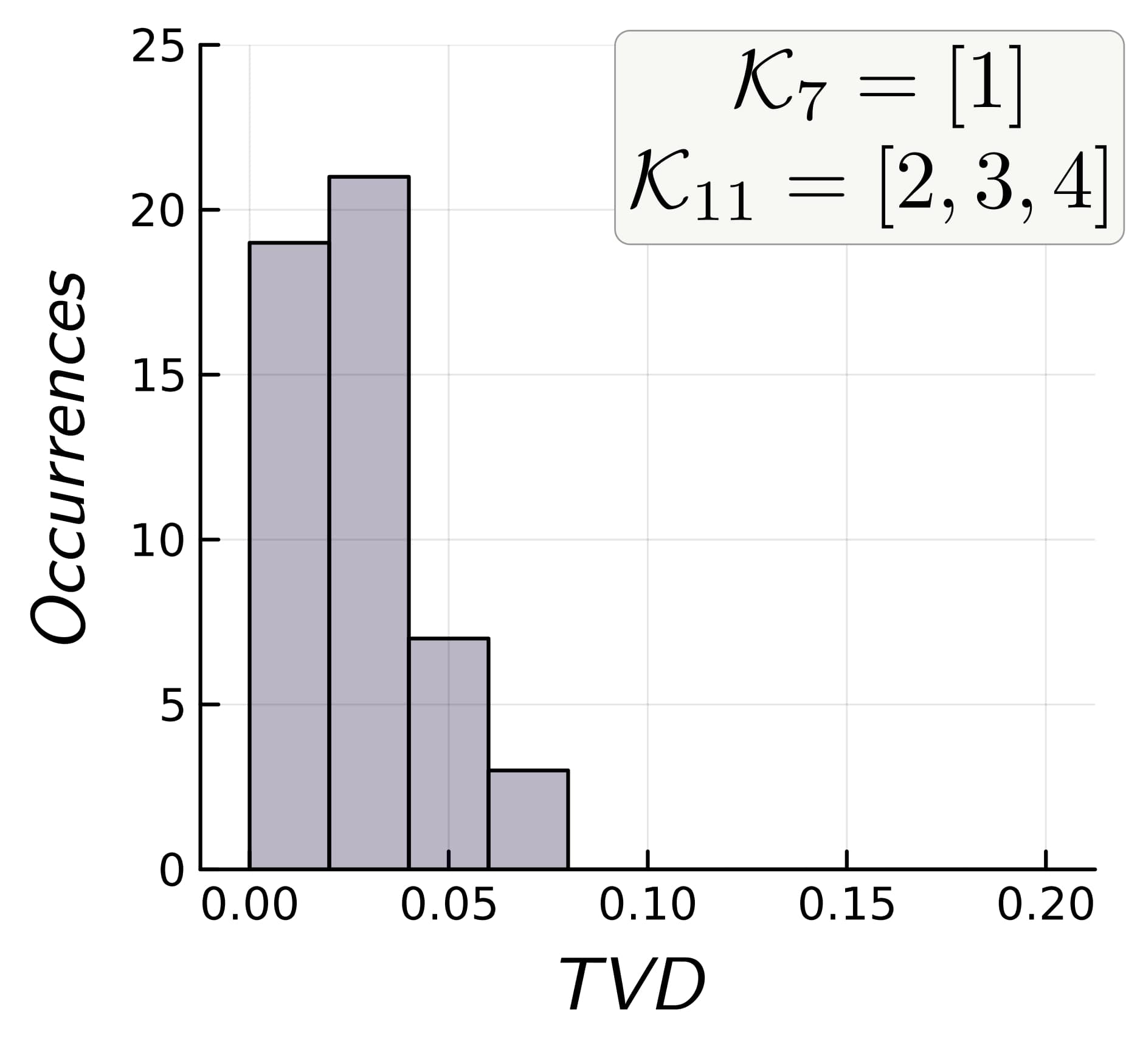}
    }
    \hfill
        \subfigure[]{
        \includegraphics[width=0.32\textwidth]{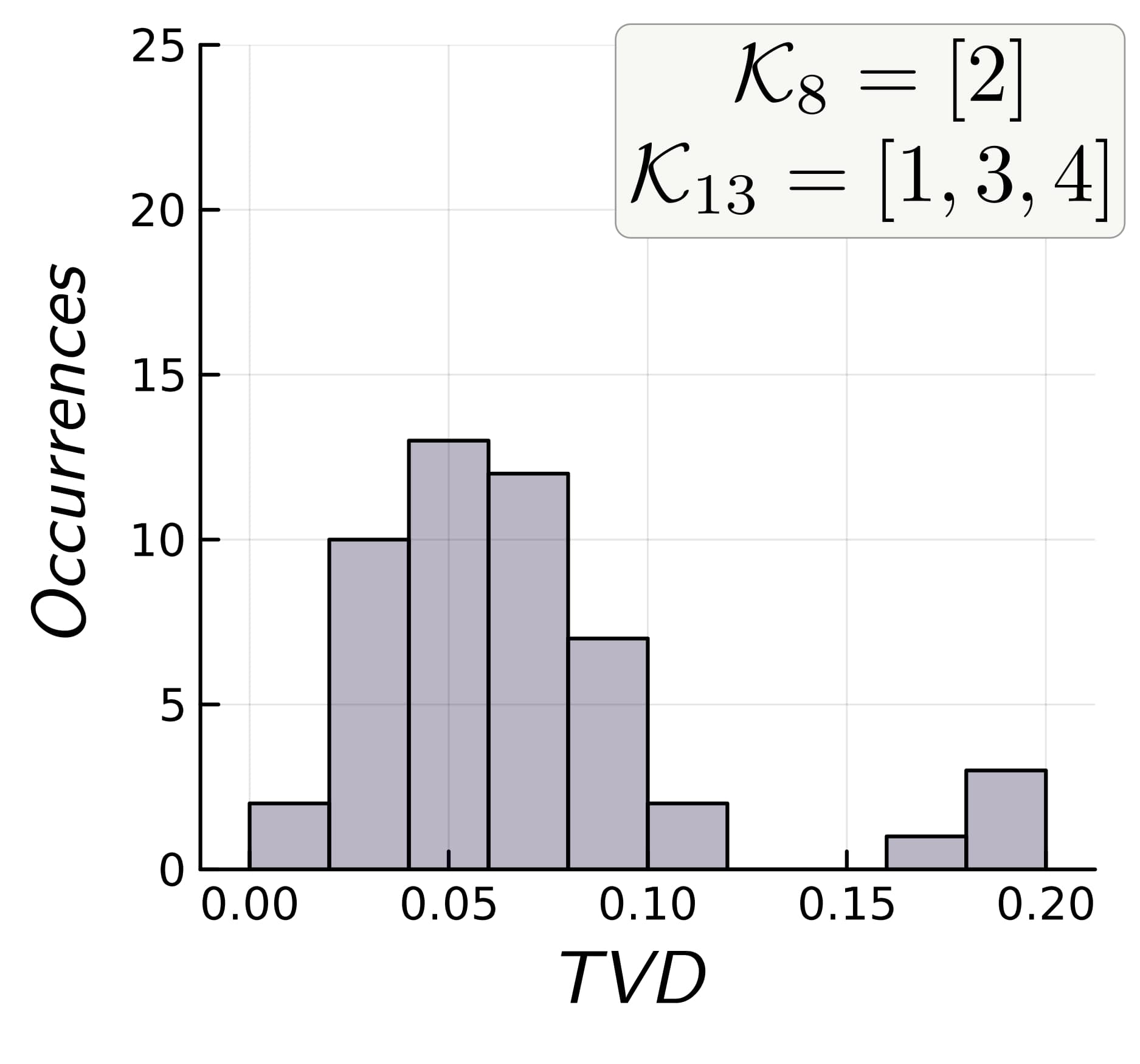}
    }
    \hfill
        \subfigure[]{
        \includegraphics[width=0.32\textwidth]{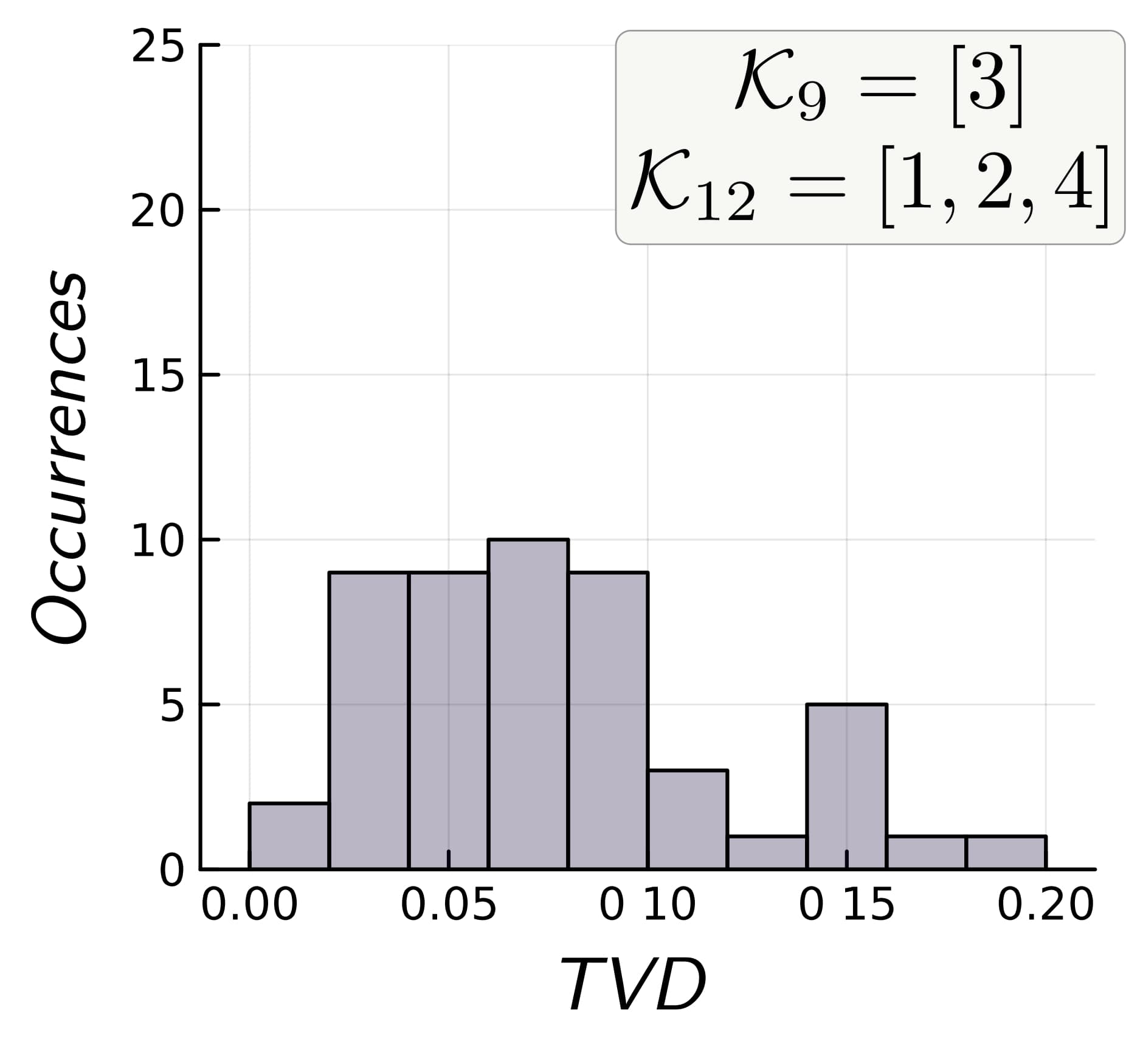}
    }
    \hfill
        \subfigure[]{
        \includegraphics[width=0.32\textwidth]{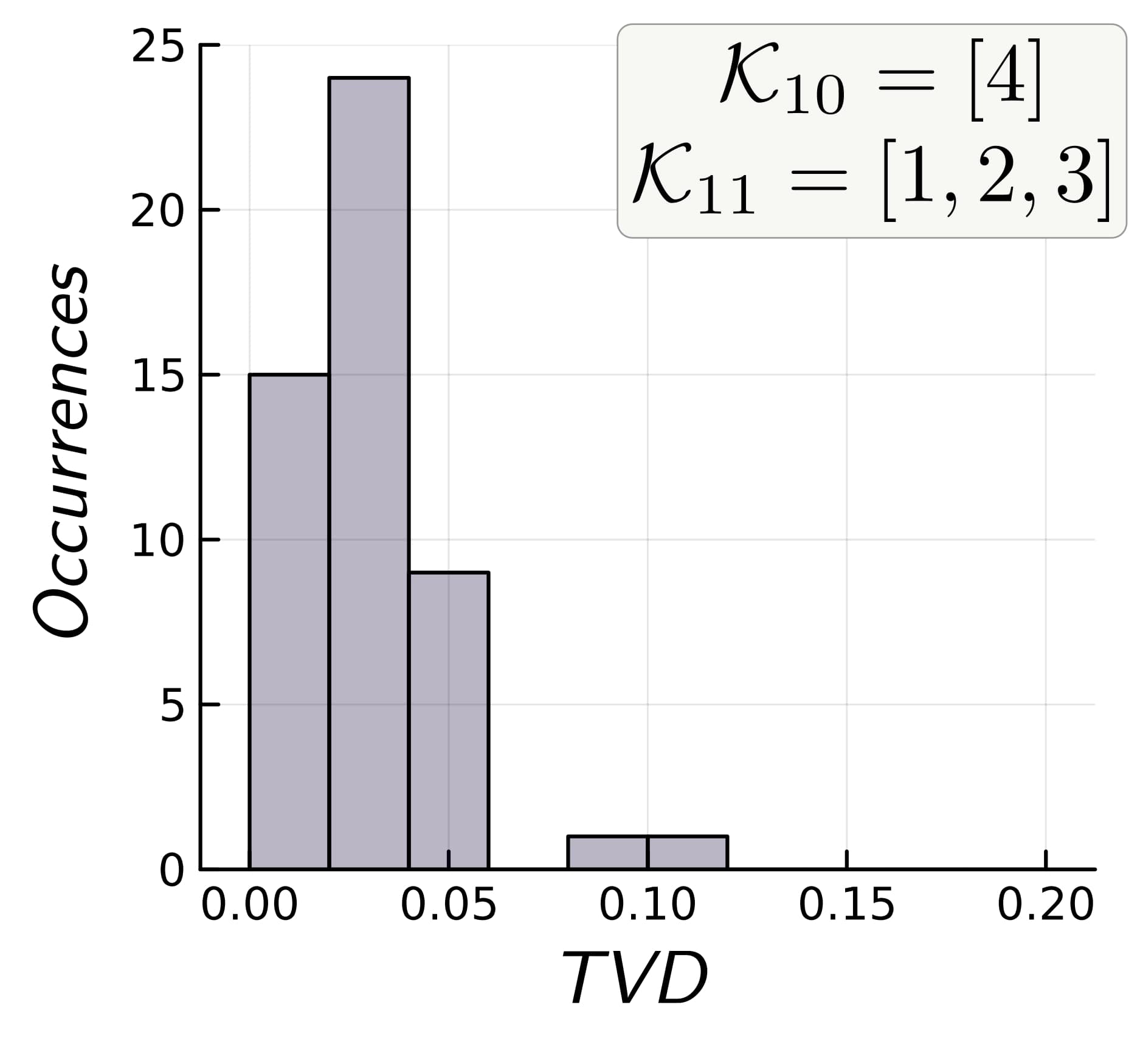}
    }
    \caption{Histogram of the Total Variation Distance (TVD) between binned-mode probability distributions for experimental data at a near-bosonic case ($\bar{x}_q\sim 0.993$) and fully bosonic ($\bar{x}_q=1$) simulations, for the 50 sampled Haar-random unitaries. \textbf{(a-g)} all partitions shown (symmetric partitions are shown together).}
    \label{fig:tvd_near_bos}
\end{figure*}

Fig.~\ref{fig:tvd_near_bos} shows for all possible choices of single-mode and two-mode bins, an histogram of the TVDs between the ideal and experimental scenario using data from the implemented ensemble of 50 Haar-random interferometers. Analysing these distributions can potentially be useful to achieve a better understanding about which unitaries are implemented less accurately. Such an analysis may help guide future improvements at the hardware level.   

\section{Single unitary $P(k)$ - non-monotonic behaviour}\label{sec:app-p(k)}

In this section we report the results for the binned-mode photon-number probability distributions for all partial distinguishability configurations measured (as opposed to only the highest indistinguishability shown in the main text). 
\\

\begin{figure*}[t]
    \centering
    \subfigure[]{
        \includegraphics[width=0.32\textwidth]{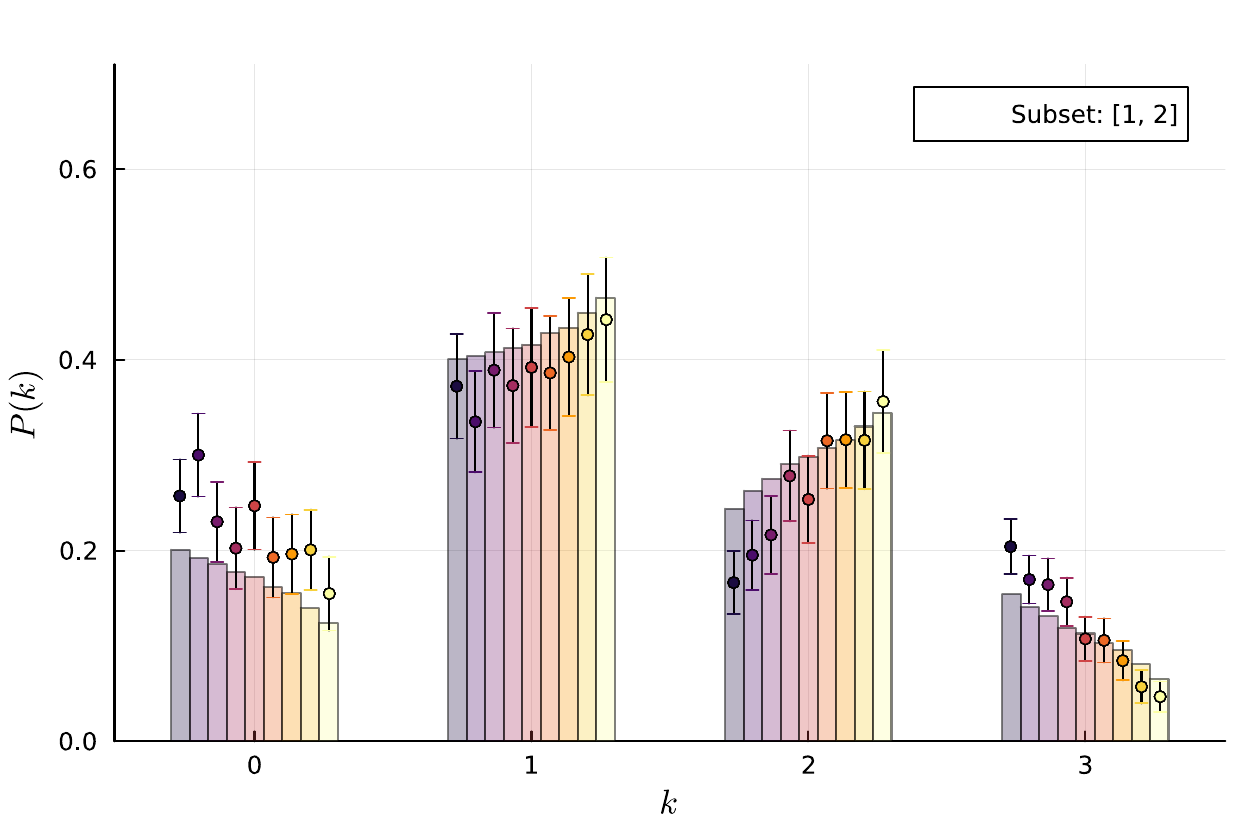}
    }
    \hfill
    \subfigure[]{
        \includegraphics[width=0.32\textwidth]{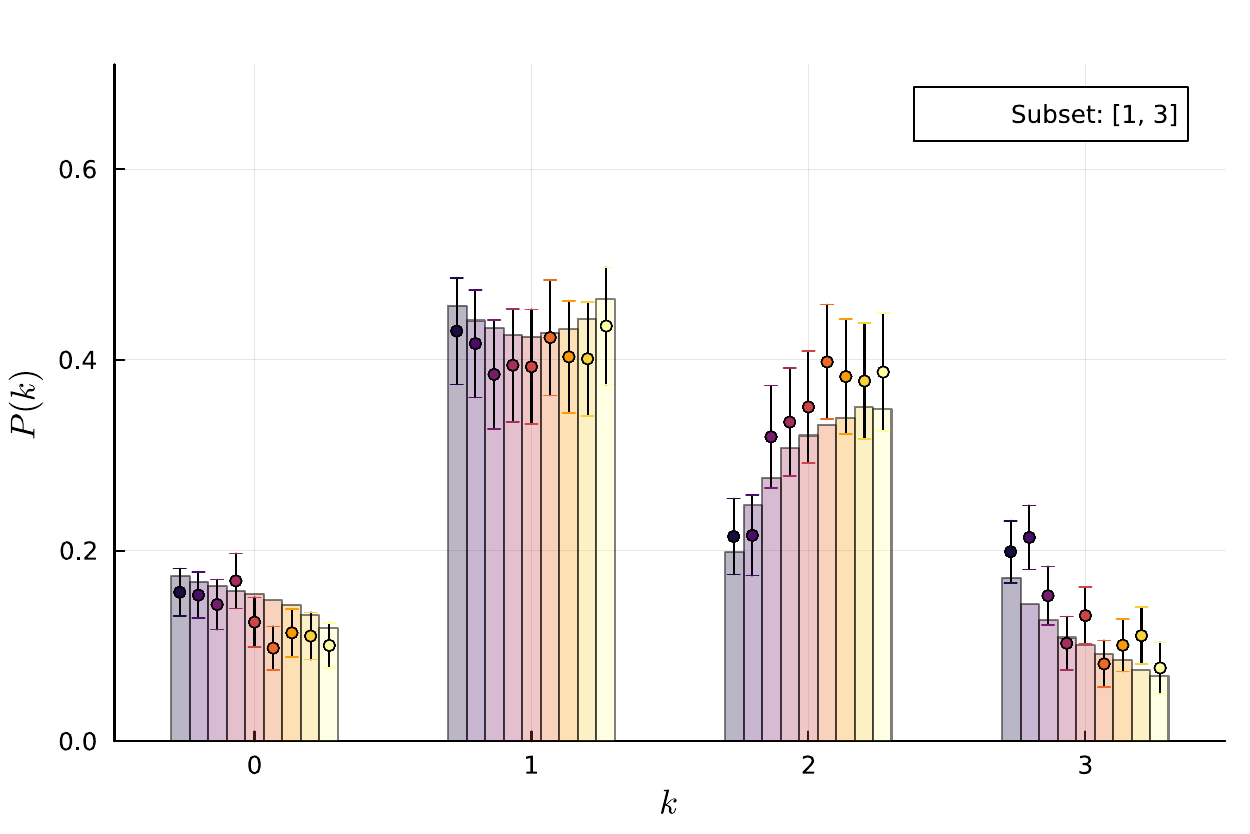}
    }
    \hfill
        \subfigure[]{
        \includegraphics[width=0.32\textwidth]{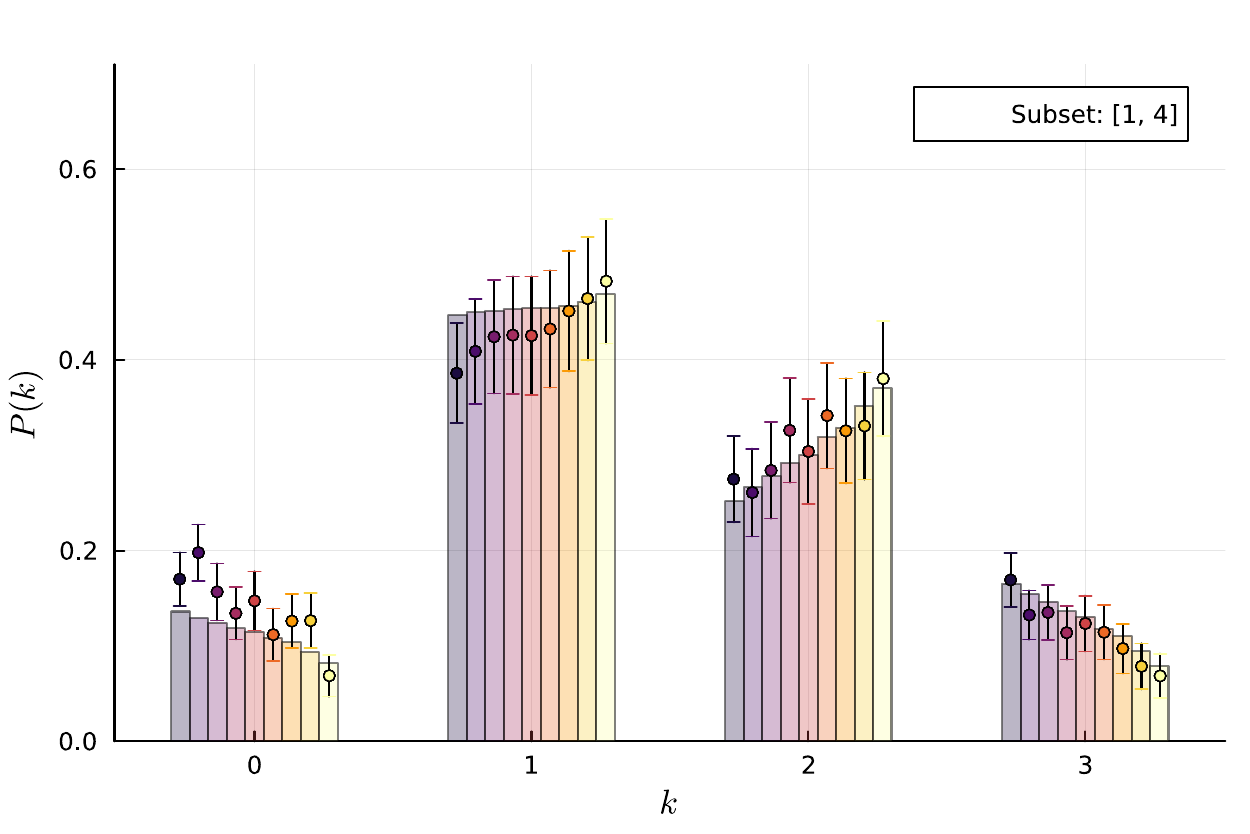}
    }
    \hfill
        \subfigure[]{
        \includegraphics[width=0.32\textwidth]{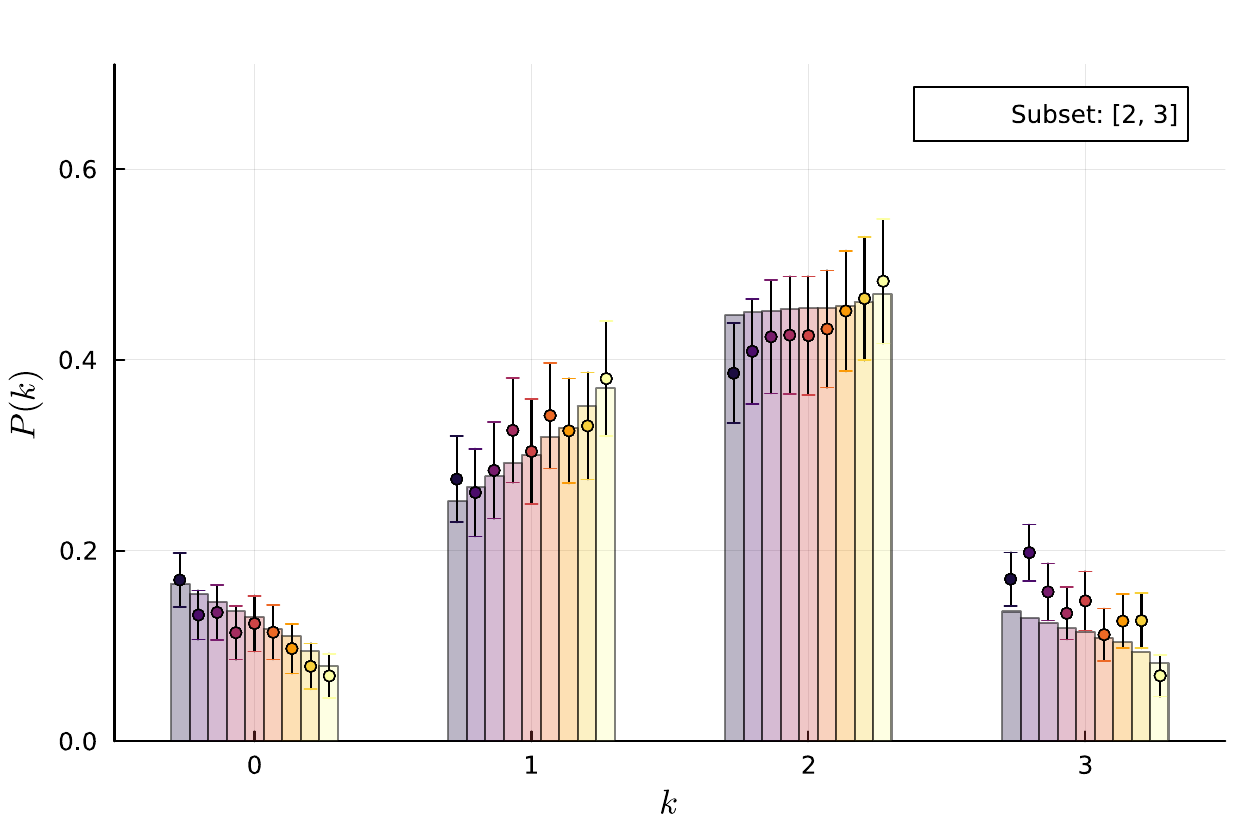}
    }
    \hfill
        \subfigure[]{
        \includegraphics[width=0.32\textwidth]{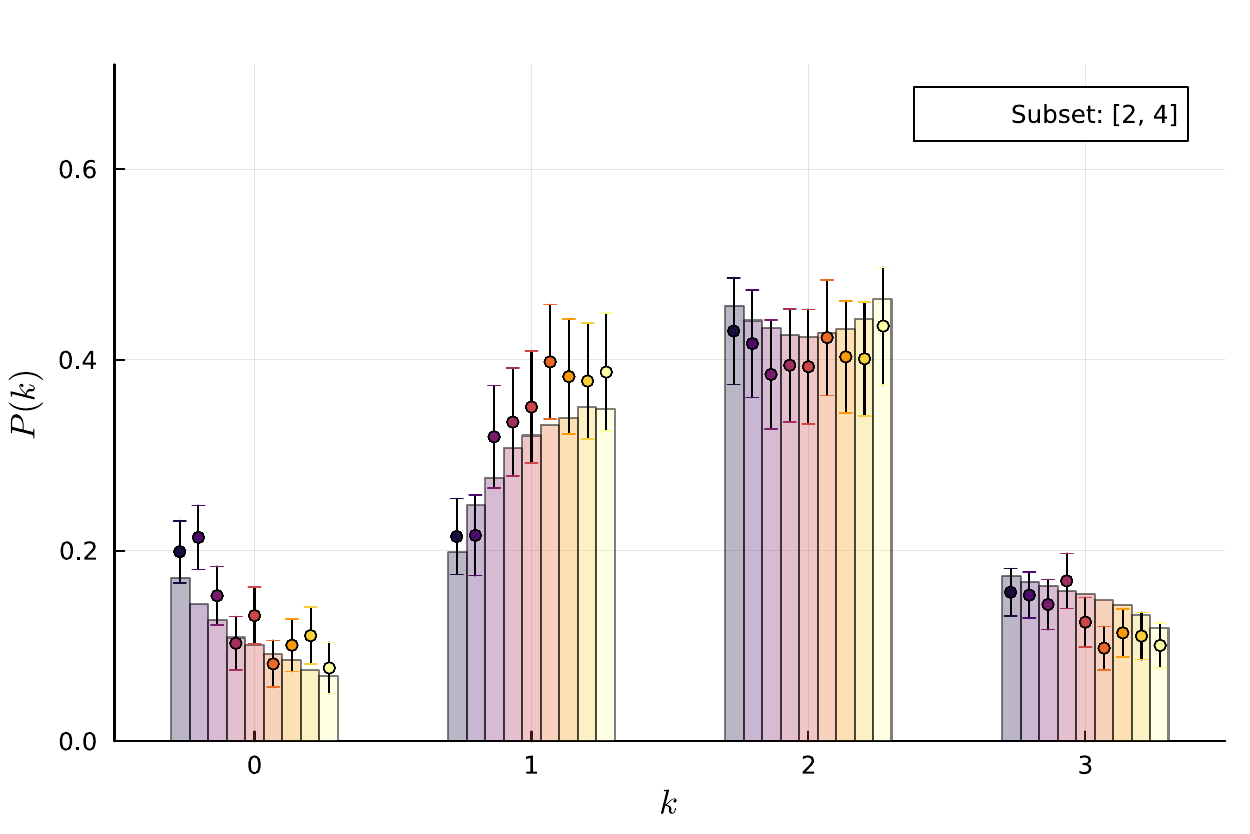}
    }
    \hfill
        \subfigure[]{
        \includegraphics[width=0.32\textwidth]{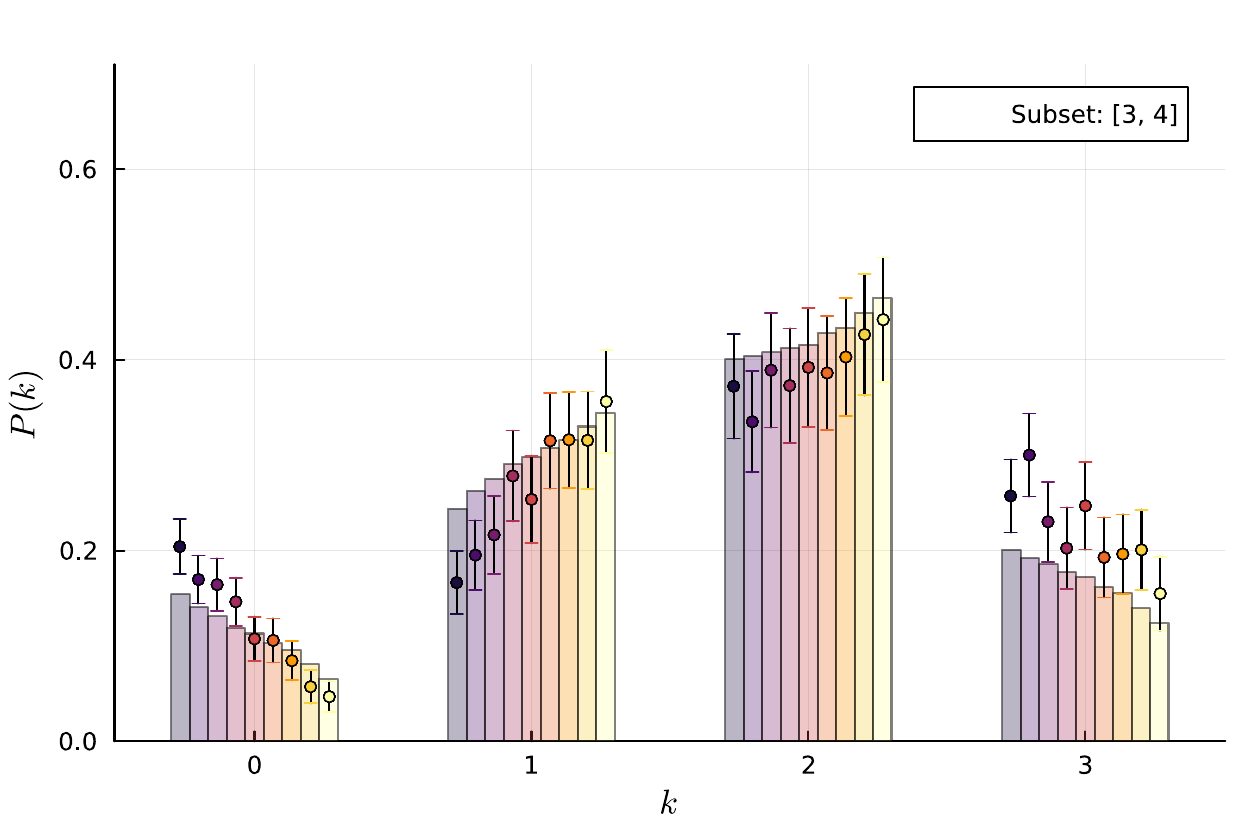}
    }
    \hfill
        \subfigure[]{
        \includegraphics[width=0.32\textwidth]{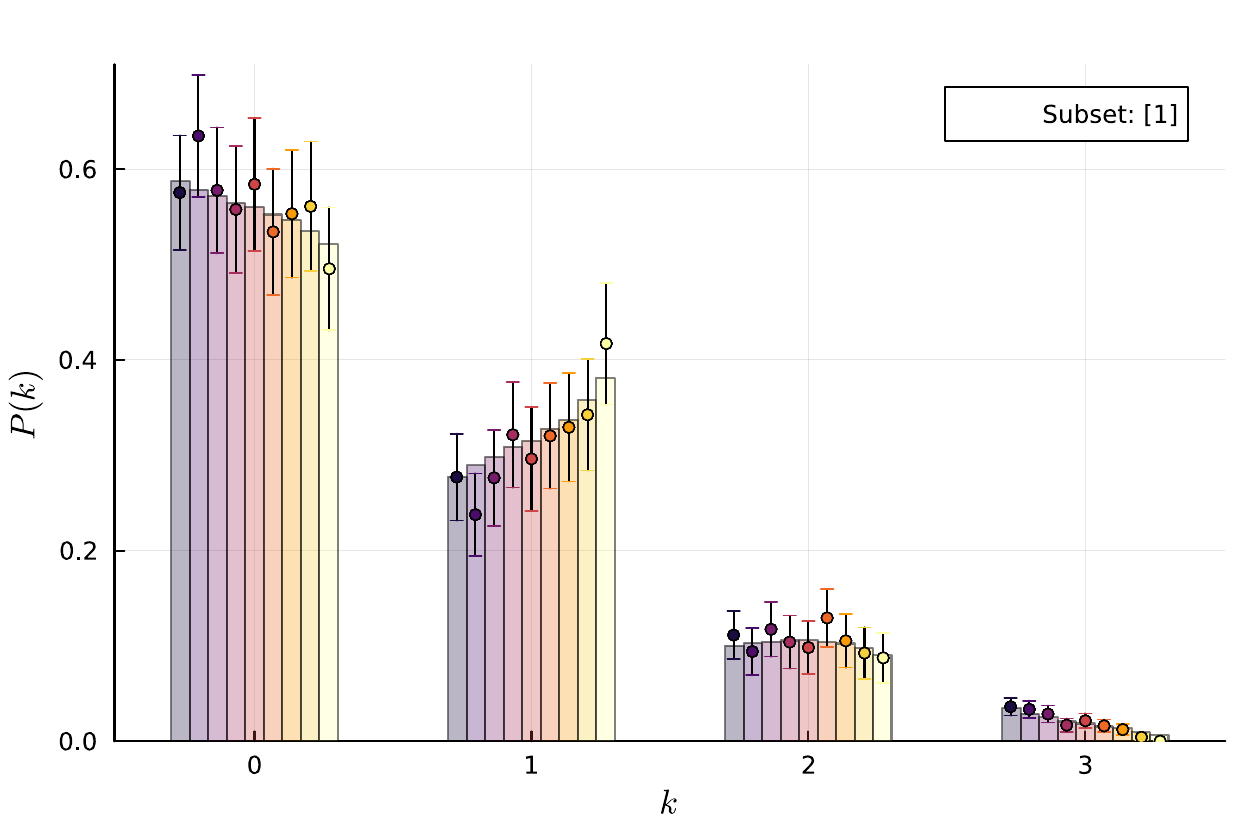}
    }
    \hfill
        \subfigure[]{
        \includegraphics[width=0.32\textwidth]{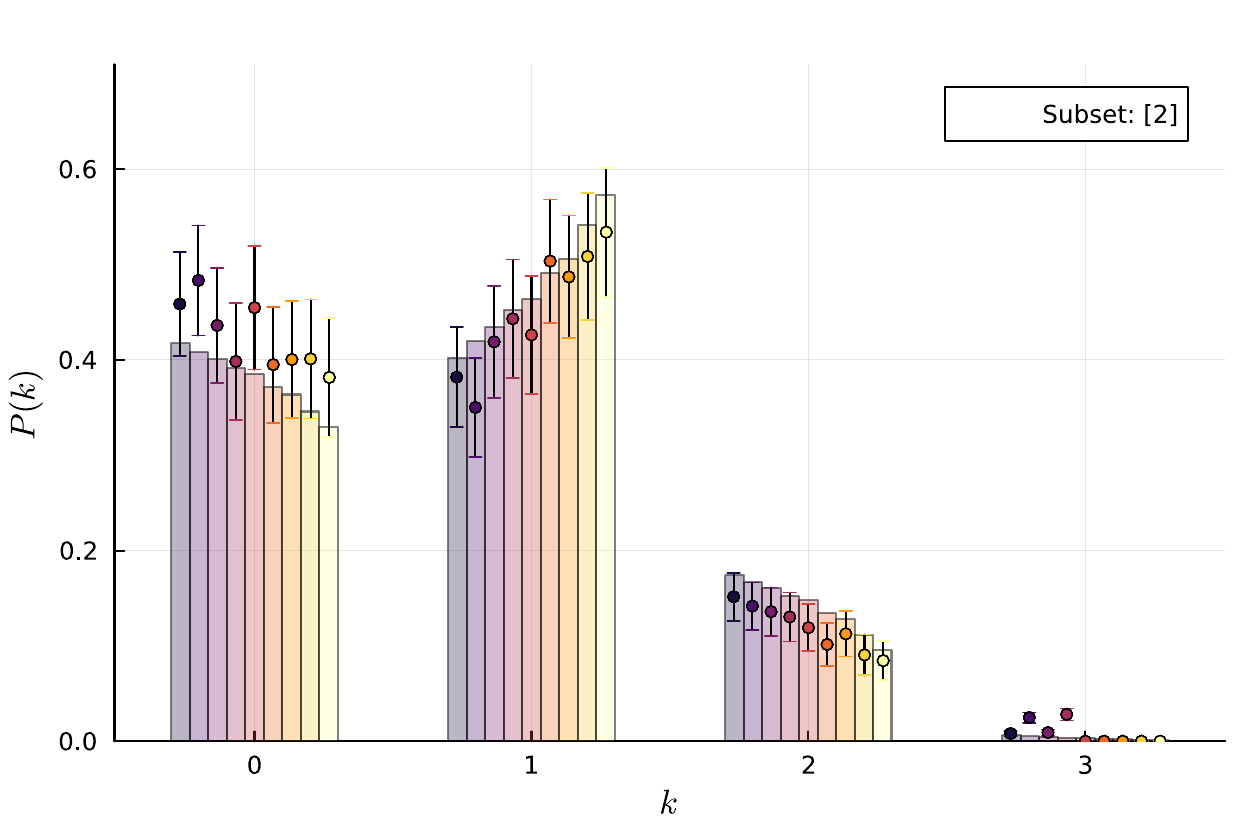}
    }
    \hfill
        \subfigure[]{
        \includegraphics[width=0.32\textwidth]{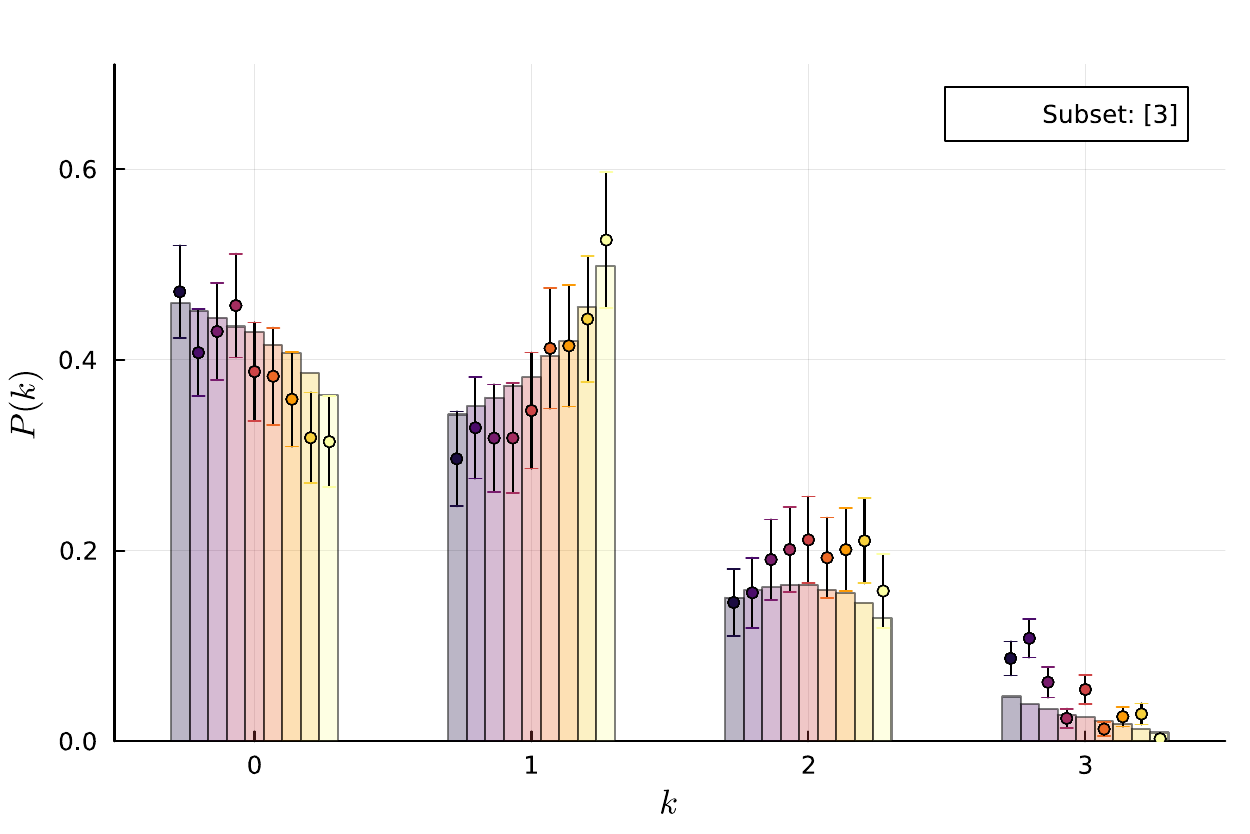}
    }
    \hfill
        \subfigure[]{
        \includegraphics[width=0.32\textwidth]{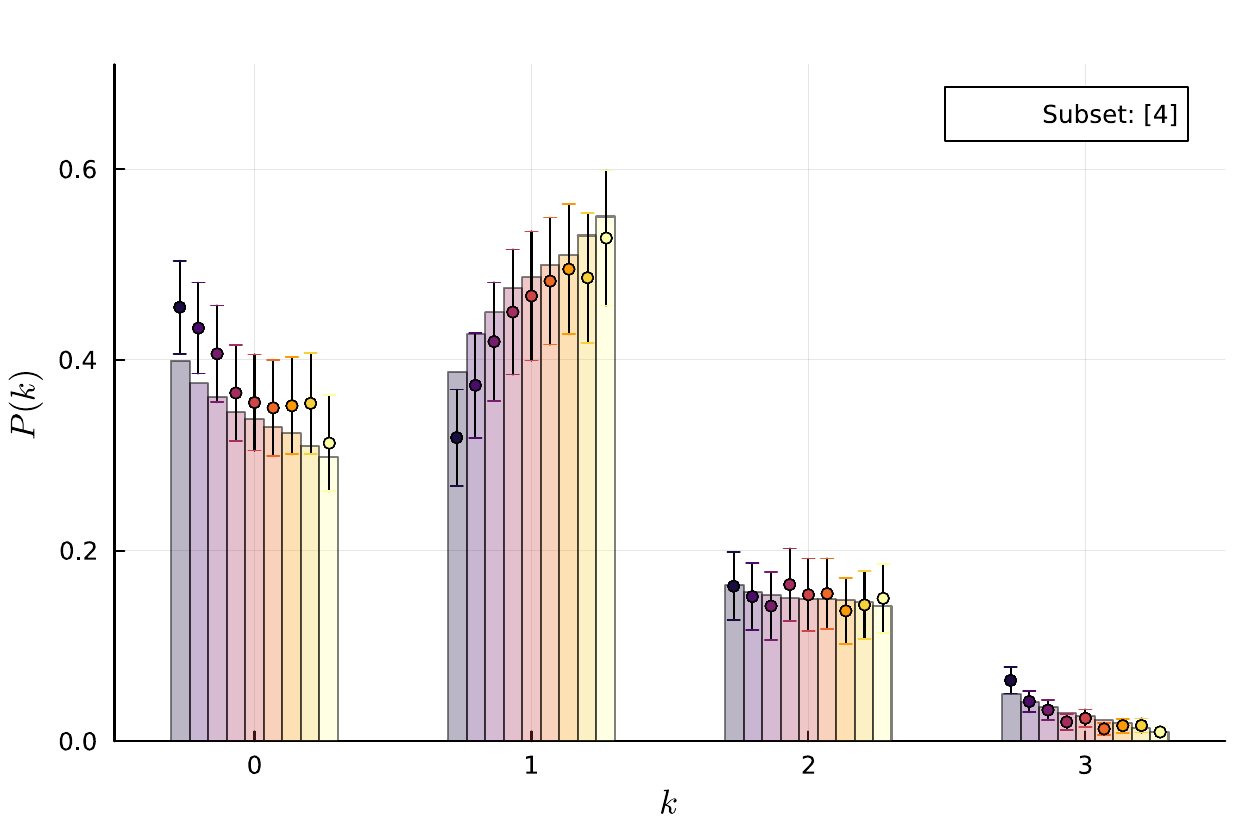}
    }
    \hfill
        \subfigure[]{
        \includegraphics[width=0.32\textwidth]{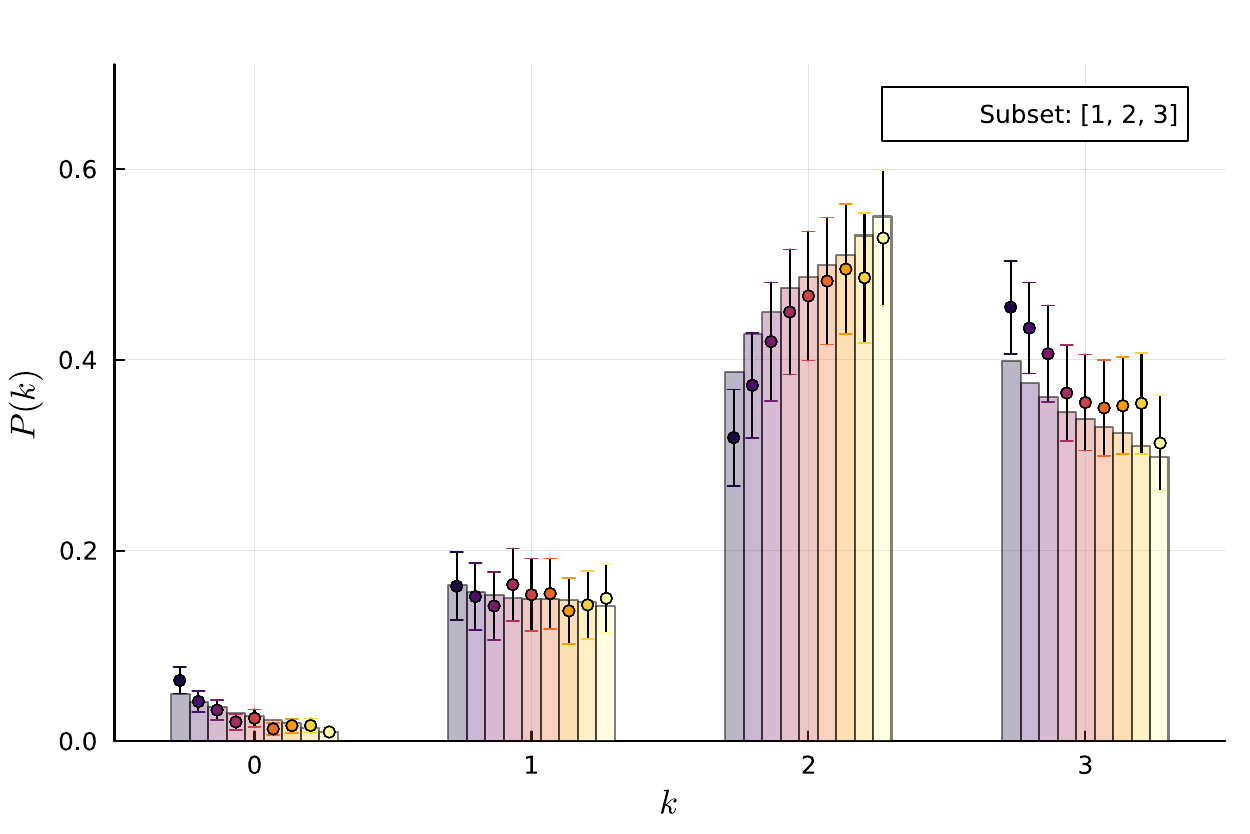}
    }
    \hfill
        \subfigure[]{
        \includegraphics[width=0.32\textwidth]{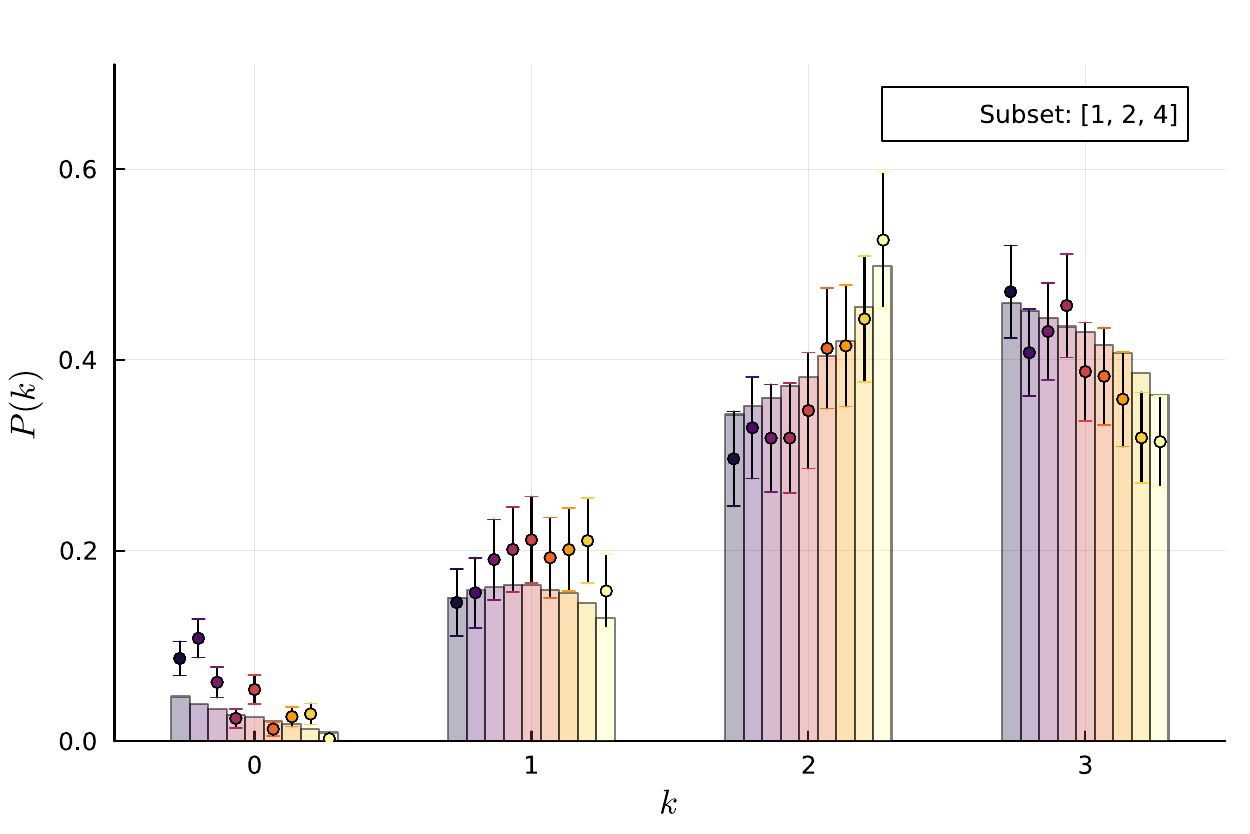}
    }
    \hfill
        \subfigure[]{
        \includegraphics[width=0.32\textwidth]{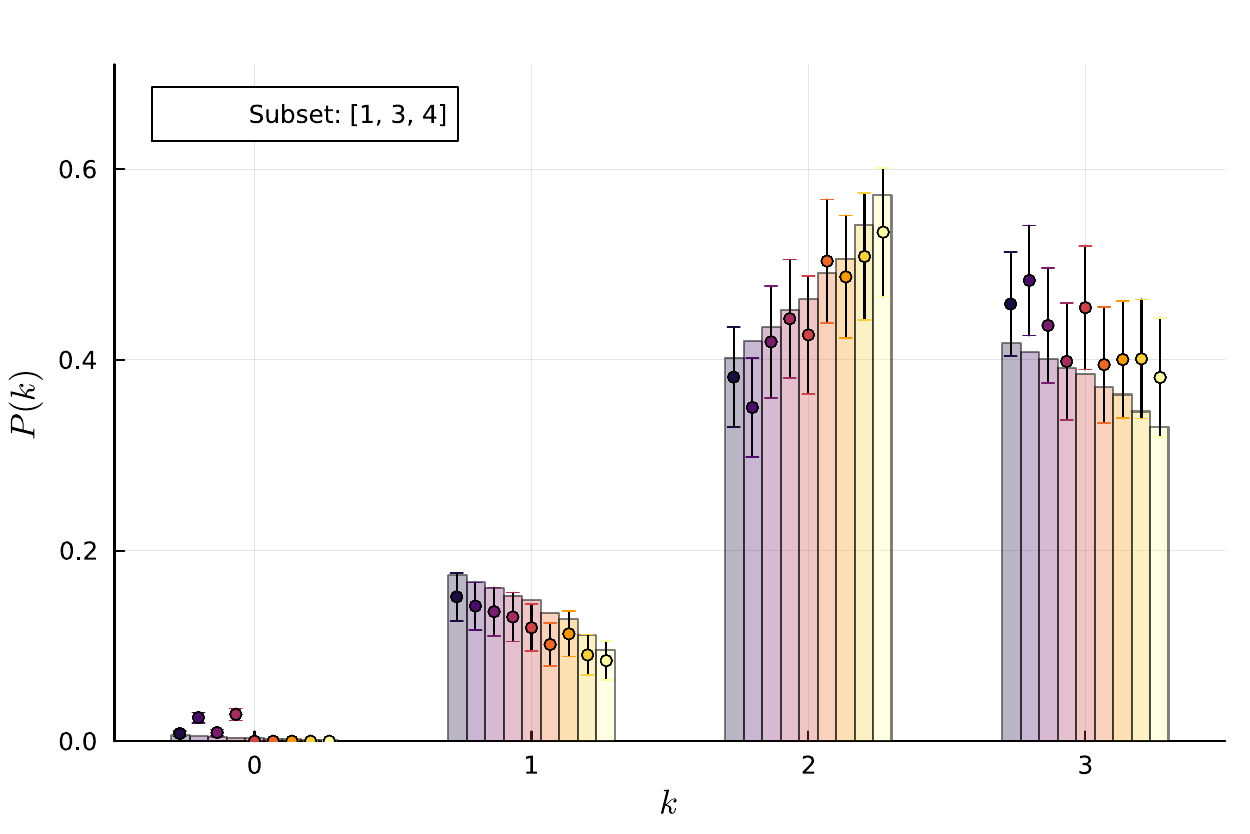}
    }
        \subfigure[]{
        \includegraphics[width=0.32\textwidth]{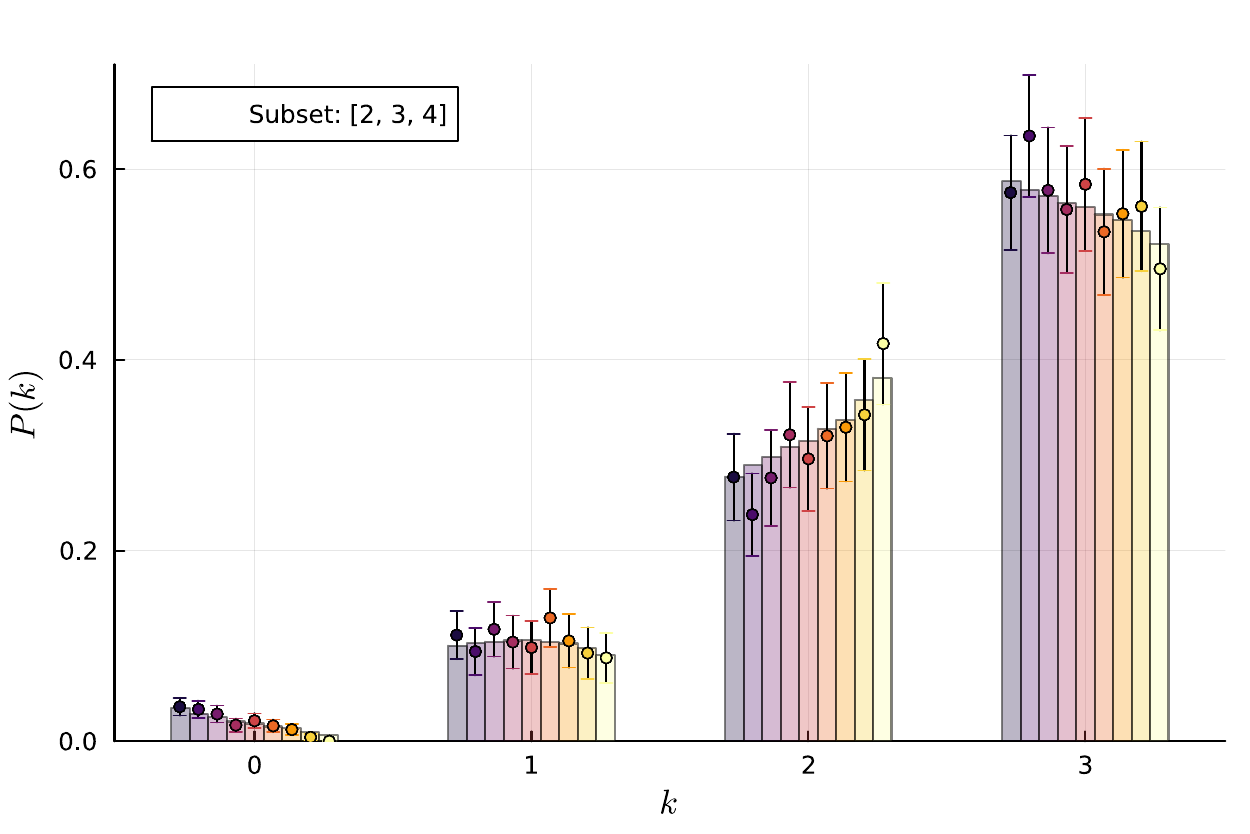}
    }
    \hfill
    \caption{Probability distributions $P(k)$ for detecting $k$ photons in a given subset of modes for the first sampled unitary $U_H^1$, across all levels of partial distinguishability studied. Black circles represent experimental data, while bars indicate the corresponding numerical simulations, ordered from highest to lowest indistinguishability (left to right). Color saturation reflects the level of indistinguishability. \textbf{(a-n)} Results are reported for all subsets.}
    \label{fig:p(k)_firstU}
\end{figure*}

Fig. \ref{fig:p(k)_firstU} shows the binned-mode photon-number probability distributions $P(\bf{k})$ for unitary $U_H^1$, for all mode binning ${\mathcal{K}}$. Circles represent experimental data, while coloured bars represent numerical simulations, which for each $\bf{k}$ group are ordered from left-to-right with increasing partial distinguishability represented by the colour. Error bars correspond to statistical noise in estimating the probability distributions via sampling. These same probability distributions for all other 49 Haar-random unitaries can be produced upon request and sent to anyone interested.\\

\textit{Non-monotonic behaviour  -  }
We would like to pay a special attention to a rare behaviour which is when the probability of observing $k$ photons in a mode (or more generally, a set of modes) $P(\bf{k})$, does not behave monotonically, that is, either strictly increasing or decreasing with partial distinguishability. The conditions for such an event to occur where first described by Tichy et al. \cite{tichy2011four} and have sparked a conversation about the strict conditions necessary for these phenomena , leading to further theoretical work \cite{ra2013nonmonotonic,bjork_non-monotonic_2014, ra_comment_2014, tichy2015_partial_distinguishability, tichy_double-fock_2015}, as well as experimental observations of non-monotonic behaviour \cite{spring_chip-based_2017, rodari_experimental_2024, menssen2017distinguishability}. In particular, it was debated whether such effects only arise with particle numbers higher than $n=5$ photons, but we have found examples in our system at $n=3$.

In this work, non-monotic behaviour can be observed in Fig. \ref{fig:p(k)_firstU}.(b) $\mathcal{K} = [1,3]$ for $k=1$ (and its symmetric counterpart $\mathcal{K} = [2,4]$ for $k=2$), as well as fig. \ref{fig:p(k)_firstU}.(i) $\mathcal{K} = [3]$ for $k=2$ (and its symmetric counterpart $\mathcal{K} = [1,2,4]$ for $k=1$).

\section{Sensitivity phase errors in the interferometer}\label{app:phase}

In this section we discuss how errors in the phases appearing in the unitary $U$ do not affect binned distributions for single-mode bins, while for bins of size two or larger they can contribute to deviations between theoretically computed values and experimentally measured ones. 

A simple argument to show the photon number distribution of a  single output mode $k$ can only depend on $|U_{kj}|$ is the following: any addition of local phase shifters at the input or output of the interferometer $U$ does not affect boson sampling outcome probabilities. We can use this freedom to put all the phases of the $k$th row of the unitary to zero, i.e. after this transformation, the matrix elements of the $k$th row will be given by $|U_{kj}|$.

\FloatBarrier


Let us now consider the case of a bin of two modes, labeled as $k_{1},k_{2}$. The $H$ matrix, which appears in the characteristic function of the binned distribution,  takes the form:
\begin{equation}
    H_{ij}=U_{k_{1}i}^{*}U_{k_{1}j}+U_{k_{2}i}^{*}U_{k_{2}j}, 
\end{equation}
which, in general, are complex valued. The argument used above cannot be applied to this instance of the problem, since the complex part appears from relative complex phases between two rows, which cannot be always sistematically corrected. As a consequence, it can be seen that the coefficients $c_j$ (except for $c_1$) will depend on the complex phases of the unitary matrix.


To see this let us consider  two different unitaries $U$ and $\Tilde{U}$, such that $|U_{i,j}|=|\tilde{U}_{i,j}|$ but they may have different complex phases,  and let us denote respectively $H$ and $\Tilde{H}$ the matrix appearing in the characteristic function. Let us suppose for both the same distiguishability matrix $x_{ij}=x, \forall i\neq j$, with $x\in [0,1]$. Let us focus our attention on the coefficients $c_{a}$ appearing in Eq. (\ref{eq: coefficient GF}), and compute the difference between the two cases.
\begin{equation}
    c_{1}(H)-c_{1}(\Tilde{H})=0
\end{equation}
\begin{equation}
    c_{2}(H)-c_{2}(\Tilde{H})=x^{2}\sum_{i\neq j}{\bigg (}H_{i,j}H_{j,i}-\Tilde{H}_{i,j}\Tilde{H}_{j,i}{\bigg )}\geq 0
\end{equation}
The terms $c_{j}$ can be written as complex polynomials in the variable $x$. Given this we can use the Parseval–Plancherel identity:
\begin{equation}
    \sum_{m}|P(m)-Q(m)|^{2}=\int dy |\hat{P}(y)-\hat{Q}(y)|^{2}
\end{equation}
where $P,Q$ are two probability distribution and $\hat{P},\hat{Q}$ are their Fourier transform. This implies that a two (and more) modes bin has a signature of the errors in the phases of the elements composing the unitary $U$. On top of that we can notice that this effect is not present with fully distinguishable particles $x=0$, which suggests that the higher is the indistinguishability the larger the distance between the two cases.

\section{GBP with noisy matrices} \label{sec:app-NoisyMatrixGBP}

In this section, we report the results of simulations involving noisy interferometers, aimed at accounting for discrepancies between the experimental results and "clean" simulations (i.e., those considering only the controlled partial distinguishability error between photons). When dealing with programmable linear interferometers, there is always some error in dialing the targeted unitary transformation onto the device. This discrepancy can be quantified using the \textit{Matrix Amplitude Fidelity} $\mathcal{F}_A$ between the targeted ("set") and actual ("get") unitary matrices that describe the interferometer:
\begin{equation} 
\mathcal{F}_A = \frac{1}{M}\text{Tr}(|U_\text{set}|^\dagger \cdot|U_\text{get}|), 
\end{equation}
where $M$ is the number of modes in the interferometer, $|U_\text{set}|$ and $|U_\text{get}|$ are the element-wise amplitude values of the targeted and actual unitary matrices, respectively, and $\cdot$ denotes standard matrix multiplication. Here, we test the hypothesis that this noise in programming the unitary is responsible for the experimental points in Fig.~\ref{fig:Fig4genBunching} dipping below 0.

Ideally, one could access the actual matrix implemented on the chip during the experiments. Matrix characterization protocols, such as those described in \cite{laing_super-stable_2012}, provide a method to achieve this by directly characterizing the "get" matrix and then simulating its effects while excluding any other sources of noise. However, in practice, these protocols require adaptive measurements and long integration times, which are incompatible with the experimental procedures used in this work. As a result, we are compelled to adopt an alternative approach to test the hypothesis.

\begin{figure}[H]
    \centering
    \includegraphics[width=0.98\linewidth]{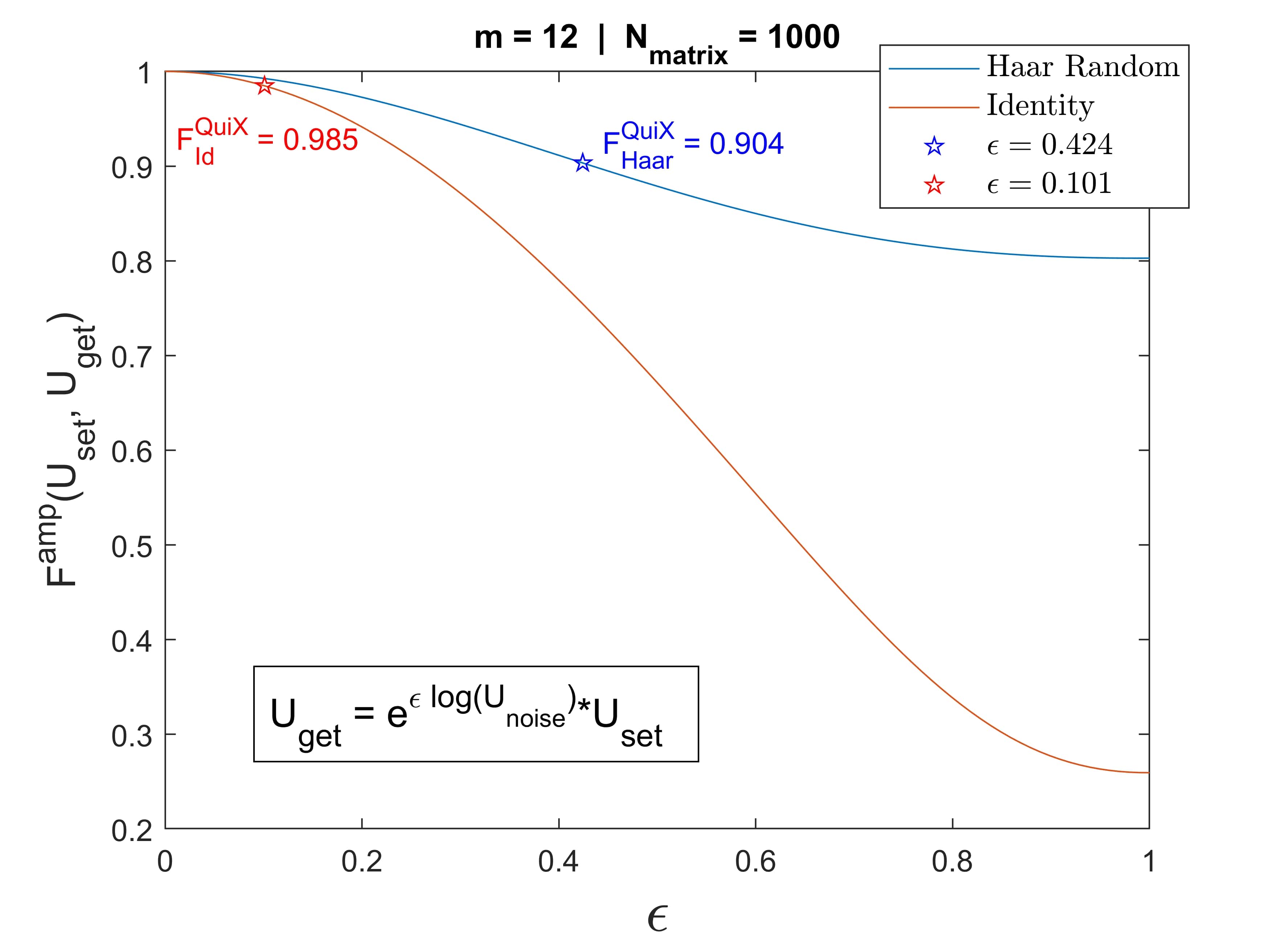}
    \caption{The Matrix Amplitude Fidelity, $\mathcal{F}_A$, between the noisy matrix $U_\text{get}$ and the targeted unitary $U_\text{set}$ of size $m=12$ as a function of the noise strength parameter $\epsilon$. The noisy matrix is generated using the model $U_\text{get} = e^{\epsilon \log(U_\text{noise})} \cdot U_\text{set}$, where $U_\text{noise}$ is a random unitary drawn from the Haar measure. Two cases are shown: one where the target interferometer $U_\text{set}$ is the identity matrix (red curve) and another where $U_\text{set}$ is a Haar-random unitary (blue curve). The Haar-random curves are generated by averaging the results of $N_\text{Matrix} = 1000$ Haar-random unitaries per $\epsilon$ value. The stars on the curves mark the $\epsilon$ values corresponding to the amplitude fidelities reported by QuiX Quantum. The inset equation defines the noisy matrix model used in the simulations.}
    \label{fig:app-noisyMatrixFid}
\end{figure}

To test this hypothesis, we simulate noisy matrices by adding random noise to the targeted unitary transformations, with the noise strength controlled by a parameter $\epsilon$. The noise model is defined as:
\begin{equation} U_\text{target}^\text{noisy}(U_\text{target},\epsilon) = U_H^\epsilon\cdot U_\text{target} = e^{\epsilon\log(\tilde{U}_{H})}\cdot U_\text{target}, \end{equation}
where $\epsilon \in [0,1]$, $\tilde{U}_H$ is a random unitary matrix drawn from the Haar measure that acts as the noise matrix, and $\cdot$ is the standard matrix multiplication. The rationale behind this approach is that errors in programming the unitary transformations occur on average in an isotropic and homogenous manner across the physical photonic chip, making this random noise model an appropriate approximation of the physical error process.

For each level of noise strength, we compute $\mathcal{F}_A$ between the noisy matrix and the targeted unitary. By varying $\epsilon$, we generate a range of amplitude fidelity values and identify the $\epsilon$ value that most closely matches the amplitude fidelity values reported by QuiX, the manufacturer of the photonic processor used in the experiments. This ensures that our noisy matrix simulations are consistent with the level of error expected in the experimental device.

Fig. \ref{fig:app-noisyMatrixFid} shows the results of adjusting the noise strength parameter $\epsilon$ to match the specifications reported by QuiX, where $\mathcal{F}_A = 0.904$ for Haar-random interferometers. The plot illustrates how the Matrix Amplitude Fidelity, $\mathcal{F}_A$, decreases as $\epsilon$ increases for two cases: identity interferometers (red curve) and Haar-random interferometers (blue curve). For Haar-random interferometers, the noisy matrices are generated by averaging over $N_\text{Matrix} = 1000$ Haar-random unitary matrices per $\epsilon$ value, providing a robust estimate of the fidelity decay. The stars on the curves highlight the specific $\epsilon$ values that correspond to the amplitude fidelities reported by QuiX. From this point onwards, we will use $\epsilon = 0.424$ for our noisy matrix simulations.

\begin{figure}[H]
    \centering
    \includegraphics[width=0.98\linewidth]{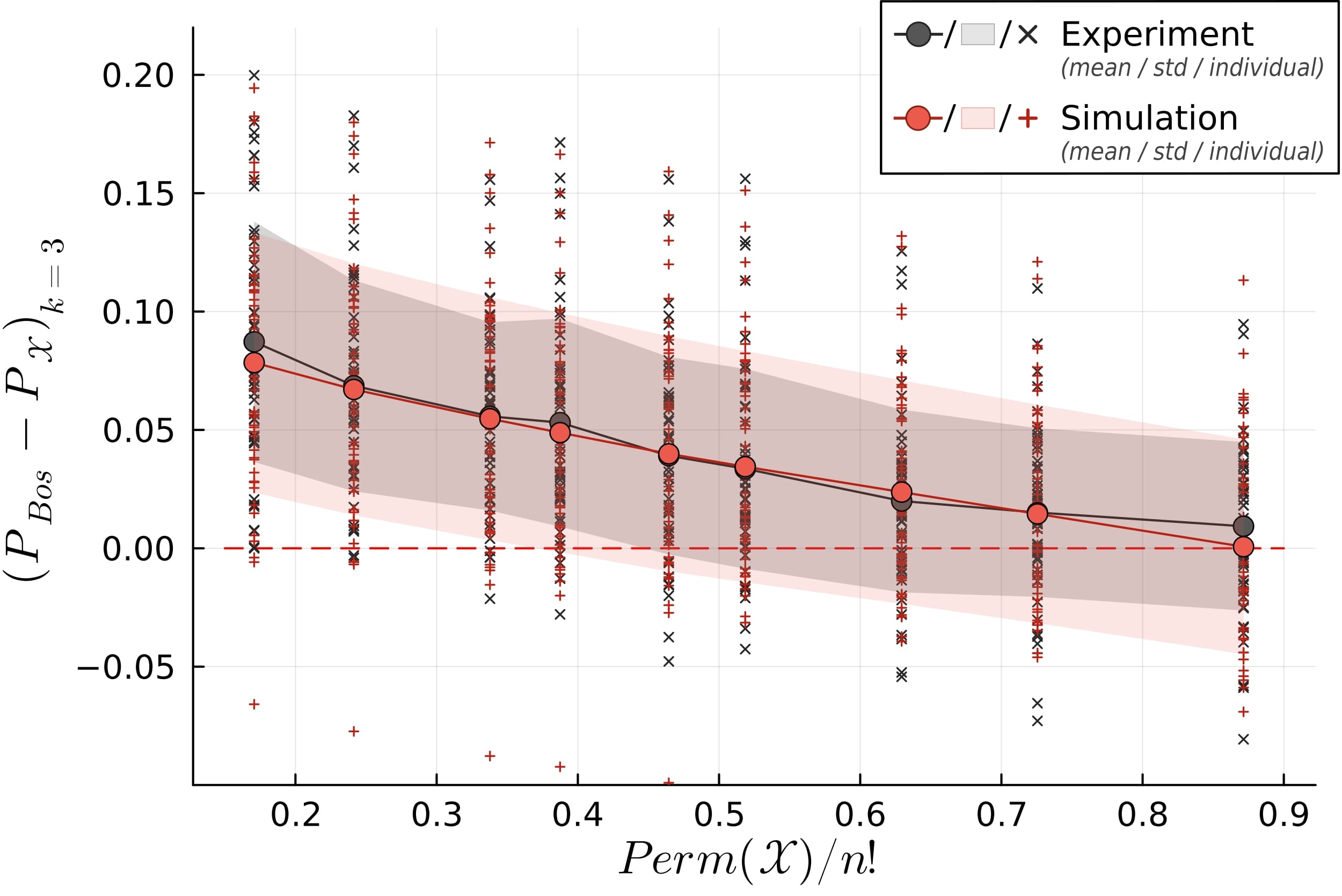}
    \caption{Difference between the generalized bunching probability (g.b.p.) at full indistinguishability ($P^\text{BOS}$, bosonic) and the g.b.p. at a given partial distinguishability ($P^\mathcal{X}$), plotted as a function of partial distinguishability ($\text{Perm}(\mathcal{X})/n!$), for all 50 random unitaries measured. Black markers represent experimental data, while red markers correspond to numerical simulations performed using the noisy matrices generated with the model described in the appendix. The solid circles/lines and shaded regions show the mean and standard deviation, respectively, for each level of distinguishability. The $\times$/+ crosses represent the data for each unitary $U_H^i$.}
    \label{fig:app-NoisyMatrixGBP}
\end{figure}

We perform simulations with noisy matrices, now generated by adding noise according to this model to the same set of 50 Haar-random unitaries studied in the main text and experiments of this work. By applying the noise model consistently across this set, we aim to directly compare the impact of noise on the results and assess how well the simulated noisy matrices replicate the experimental observations.

Fig. \ref{fig:app-NoisyMatrixGBP} shows the difference between the generalized bunching probability (g.b.p.) at full indistinguishability, $P^\text{BOS}$ (bosonic), and the g.b.p. at a given partial distinguishability, $P^\mathcal{X}$, plotted as a function of the partial distinguishability metric $\text{Perm}(\mathcal{X})/n!$. The data is presented for all 50 random unitaries studied in this work. Black markers represent the experimental results, while red markers correspond to numerical simulations performed using the noisy matrices generated with the noise model described in the appendix. These simulations incorporate the same set of 50 Haar-random unitaries as the main text, but now include additional noise to account for potential experimental imperfections in programming the unitary transformations. The solid circles and lines represent the mean values of $P^\text{BOS} - P^\mathcal{X}$ across all unitaries for each level of partial distinguishability, while the shaded regions indicate the standard deviation. The individual data points for each unitary $U_H^i$ are shown as $\times$ (experiment) and $+$ (simulation). 

The addition of noise to the simulations allows for a closer comparison with the experimental results, aiming to replicate the observed behavior, particularly the deviations in the middle and high ranges of partial distinguishability where the experimental points often fall below the 0 line. This improvement is especially apparent when compared to Fig.~\ref{fig:Fig4genBunching} in the main text, where noiseless simulations were shown to fit the experimental data much less closely, particularly in the middle range of partial distinguishability. Remarkably, these noisy simulations match the experimental data almost exactly, even without further refinement of the noise model. This agreement strongly supports the conclusion that noise in the programmed unitary transformations is the dominant source of error in the analysis. Furthermore, this reinforces the hypothesis that such noise is responsible for the observed increase in bunching, causing the experimental results to exhibit more bunching than would normally be allowed by the theoretical predictions under clean conditions.

\begin{figure}[H]
    \centering
        \includegraphics[width=0.48\textwidth]{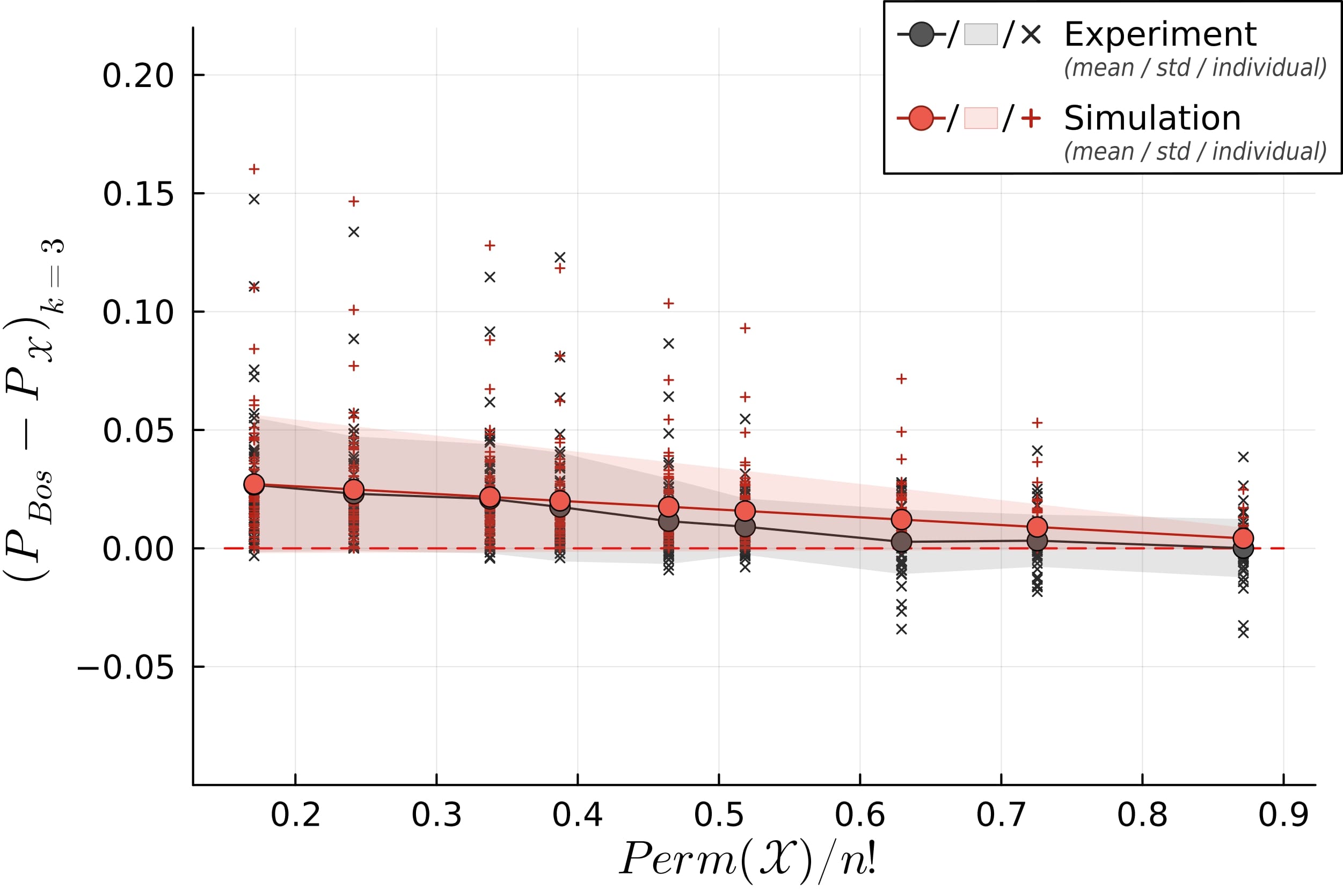}
    \caption{Difference between the generalized bunching probability (g.b.p.) at full indistinguishability ($P^\text{BOS}$, bosonic) and the g.b.p. at a given partial distinguishability ($P^\mathcal{X}$), plotted as a function of partial distinguishability ($\text{Perm}(\mathcal{X})/n!$), for all 50 random unitaries measured, for a fixed subset $\mathcal{K}_7 = [1]$.}
    \label{fig:gbp_k7}
\end{figure}

Finally, Fig.\ref{fig:gbp_k7} illustrates the differences between GBPs for a single-mode subset $\mathcal{K}_7=[1]$, which corresponds to the full bunching probability \cite{general_rules_bunching, rodari_experimental_2024}). In this particular case, the theoretical values of these probabilities are expected to be independent of the phases of the unitary matrix elements $U_{ij}$. Experimentally, this expectation is reflected in the data, as significantly fewer negative values are observed compared to other cases. This observation provides further evidence that phase mischaracterization in the elements of $U_{ij}$ is a key factor contributing to the discrepancies between the theoretical predictions and the experimental results shown in Fig.\ref{fig:Fig4genBunching}.

\bibliographystyle{unsrt}
\bibliography{references}

\end{document}